\documentclass[11pt]{article}
\usepackage[T1]{fontenc}
\usepackage[utf8]{inputenc}
\usepackage{color}
\usepackage{array}
\usepackage{booktabs}
\usepackage{url}
\usepackage{multirow}
\usepackage{amsmath}
\usepackage{amssymb}
\usepackage{graphicx}
\usepackage[authoryear]{natbib}
\usepackage[unicode=true,
 bookmarks=false,
 breaklinks=false,pdfborder={0 0 1},% backref=section,
 colorlinks=false]
 {hyperref}

\makeatletter

\providecommand{\tabularnewline}{\\}

\@ifundefined{date}{}{\date{}}
\usepackage{latexsym}
\usepackage{multirow}\usepackage{bm}%\usepackage[latin9]{inputenc}
\usepackage{url}% not crucial - just used below for the URL 
\usepackage{tabularx}
\usepackage{booktabs}
\usepackage[table]{xcolor}
\usepackage{setspace}

\usepackage{enumerate}

\newenvironment{proof}{\noindent\textit{Proof:}}{\hfill$\square$}
\usepackage{dcolumn}
\newcolumntype{.}{D{.}{.}{-1}}
\newcolumntype{d}[1]{D{.}{.}{#1}}
\usepackage{theorem}
\theoremstyle{definition}
\newtheorem{assumption}{Assumption}\newtheorem{example}{Example}\newtheorem{theorem}{Theorem}\newtheorem{remark}{Remark}
\usepackage{rotating}

\usepackage{arydshln}

\usepackage[compact]{titlesec}

\allowdisplaybreaks

\newcommand{\spacingset}[1]{\renewcommand{\baselinestretch}%
{#1}\small\normalsize}

\newcommand{\E}{\mathbb{E}}

\newcommand{\MI}{\text{MI}}
\newcommand{\DI}{\text{DI}}

\makeatother

\begin{document}
\title{\textbf{Sensitivity analysis in longitudinal clinical trials via distributional imputation}}

%\iffalse 
\author{Siyi Liu$^1$, Shu Yang$^1$, Yilong Zhang$^2$, Guanghan (Frank) Liu$^{2}$} 

\date{\vspace{-5ex}}

\maketitle
\begin{center}
$^1$Department of Statistics, North Carolina State University, Raleigh, NC, USA

$^2$Merck \& Co., Inc., Kenilworth, NJ, USA
\end{center}

%\fi 

%\thispagestyle{empty}
%\setcounter{page}{0}

\spacingset{1.5} 
\begin{abstract}
Missing data is inevitable in longitudinal clinical trials. Conventionally,
the missing at random assumption is assumed to handle missingness,
which however is unverifiable empirically. Thus, sensitivity analysis
is critically important to assess the robustness of the study conclusions
against untestable assumptions. Toward this end, regulatory agencies
often request using imputation models such as return-to-baseline,
control-based, and washout imputation. Multiple imputation is popular
in sensitivity analysis; however, it may be inefficient and result
in an unsatisfying interval estimation by Rubin's combining rule.
We propose distributional imputation (DI) in sensitivity analysis,
which imputes each missing value by samples from its target imputation
model given the observed data. Drawn on the idea of Monte Carlo integration,
the DI estimator solves the mean estimating equations of the imputed
dataset. It is fully efficient with theoretical guarantees. Moreover,
we propose weighted bootstrap to obtain a consistent variance estimator,
taking into account the variabilities due to model parameter estimation
and target parameter estimation. The finite-sample performance of
DI inference is assessed in the simulation study. We apply the proposed
framework to an antidepressant longitudinal clinical trial involving
missing data to investigate the robustness of the treatment effect.
Our proposed DI approach detects a statistically significant treatment
effect in both the primary analysis and sensitivity analysis under
certain prespecified sensitivity models in terms of the average treatment
effect, the risk difference, and the quantile treatment effect in
lower quantiles of the response, uncovering the benefit of the test drug for curing depression.

\noindent \textbf{keywords:} Longitudinal clinical trial, missing data, distributional imputation, multiple imputation, sensitivity analysis
\end{abstract}
\newpage{}

\section{Introduction}

In longitudinal clinical trials, participants are likely to deviate
from the protocol that causes the missing data. The deviations from
the protocol may include poor compliance with the treatment or loss
of follow-ups. \citet{rubin1976inference} develops a framework to
handle missingness in data. Three missing mechanisms have been proposed
as missing completely at random (MCAR), missing at random (MAR), and
missing not at random (MNAR). Missingness that is
not related to any components of the data, e.g., participants dropping
out of the trial due to work or family considerations, is categorized
as MCAR. While in most clinical studies involving patients with
missing outcomes, it is likely that the missingness depends on the
health status of patients.  For example, individuals with severe
outcomes are more likely to drop out from the study or switch to certain
rescue therapies. MAR is typically used in longitudinal
clinical trials targeting the primary analysis, which assumes that the
conditional outcome distribution stays the same between the participants
who remain in the study and the ones who drop out, i.e., the participants are assumed to take the assigned treatment even after the occurrence of missingness.
However, the MAR assumption is not verifiable and may be violated
for some drugs with a short half-life, where the treatment effect quickly fades
away once the individuals discontinue from  the active treatment, leading to a missing not at random
(MNAR) assumption. Therefore, it is vital to conduct sensitivity
analyses to explore the robustness of results to alternative MNAR-related
assumptions as recommended by the US Food and Drug Administration
(FDA) and National Research Council \citep{little2012prevention}.

The importance of defining an appropriate treatment effect estimand in the presence
of missing data has been put forward by the ICH E9(R1) working group.
Following the instructions in \citet{international2019addendum},
the estimand should give a precise description of the treatment effect
of interest from a population perspective, and account for the intercurrent
events such as the discontinuation of treatment.
% \citet{carpenter2013analysis} propose to use}\textit{\textcolor{blue}{De jure}} \textcolor{blue}{ or treatment effect estimand to answer the treatment efficacy.
In the primary analysis of the treatment effect estimand, we can assume MAR under an envisioned condition that participants with the treatment
discontinuation still follow the assigned therapy throughout the study \citep{carpenter2013analysis}.
For sensitivity or supplemental analyses, we evaluate the treatment effect under scenarios that deviate from MAR and call these settings sensitivity analyses for simplicity throughout the paper.
% consider scenarios that deviate from the MAR assumption and evaluate the treatment effect } Although these analyses may be targeted for different estimands, throughout the paper we evaluate the treatment effect under these settings and call the settings as sensitivity analyses for simplicity. 

In sensitivity analyses, we consider several plausible missingness scenarios under MNAR based on the pattern-mixture model (PMM; \citealp{little1993pattern}) framework, which we call the ``sensitivity models''. Our main focus in this paper is on the
jump-to-reference (J2R) scenario proposed by \citet{carpenter2013analysis},
which assumes that the missing outcomes in both treatment groups will
have the same distributional profile as those in the control group
with the same covariates. We also briefly
introduce other sensitivity models such as return-to-baseline (RTB)
and washout imputation, which have been used in the FDA statistical
review and evaluation reports for certain treatments (e.g., \citealp{rtb2016tresiba}).
% \textcolor{blue}{
% In all scenarios that involve missing data, we adopt the pattern-mixture model (PMM; \citealp{little1993pattern}) framework to define
% .}
Although we focus on specific sensitivity models, our framework can
be extended readily to other imputation mechanisms and the mixture
of imputation strategies in sensitivity analyses.

To handle missingness in sensitivity analyses, the likelihood-based
method and multiple imputation (MI) are the two most common approaches.
The likelihood-based method typically utilizes the ignorability of
the missing mechanism under MAR to draw valid maximum-likelihood inferences
given variation independence, i.e., the parameters that control the
missing mechanism and the model parameters are separable. For longitudinal
clinical trials with continuous responses, one can fit a mixed model
with repeated measures (MMRM) and incorporate the missing information
to obtain inferences (e.g., \citealp{mehrotra2017missing}; \citealp{zhang2020likelihood}).
While it is efficient, the analytical form for the likelihood-based
method is only feasible to derive under restrictive scenarios such
as normality or when we are dealing with mean types of estimands,
and it requires rederivations if the missingness pattern changes.
MI developed by \citet{rubin2004multiple} resorts to using computational
techniques to ease the analytical requirements from the likelihood-based
method. The FDA and National Research Council \citep{little2012prevention}
highly recommend the use of MI and Rubin's MI combining rules to get
inferences due to its flexibility and simplicity. However, \citet{wang1998large}
reveal that the MI estimator is not efficient in general. Moreover,
the inefficiency of MI can be more severe in terms of interval estimation,
where the variance estimation using Rubin's rule may not be consistent
even when the imputation and analysis models are the same correctly
specified \citep{robins2000inference}. In sensitivity analyses, overestimation
of the variance using Rubin's rule is commonly detected in literature
(e.g., \citealp{lu2014analytic}; \citealp{liu2016analysis}; \citealp{yang2016note}).
The motivating example in Section \ref{sec:example}
further shows an alteration of the study conclusion due to the conservative
variance estimator, where the same statistically significant treatment
effect fails to be detected in the sensitivity analysis, rising a
dilemma for the investigators in the process of decision-making.
To overcome the problem, the variance estimation derived from the bootstrap approach is applied. But it is more computationally
intensive than the traditional Rubin's method since it requires the re-imputation
of the missing components and the reconstruction of the imputation model
per bootstrap iteration.

In this paper, we propose distributional imputation (DI) based on the
idea of Monte Carlo (MC) integration \citep{lepage1978new} and develop
a unified framework to conduct sensitivity analyses using DI in longitudinal
clinical trials. The motivation of DI is to impute the missing components
from the target imputation model given the observed data and use the
mean estimating equations approximated by MC integration to draw efficient
inferences. The implementation consists of three major steps: first,
obtain the model parameter estimator based on the observed data; second,
impute the missing values from the estimated sensitivity model; and
third, derive the DI estimator of the parameter of interest by jointly
evaluating the entire imputed dataset through mean estimating equations.
We show that the DI estimator is consistent and asymptotically normal.
We also propose a weighted bootstrap procedure for variance estimation,
which incorporates the uncertainty from model parameter estimation
and target parameter estimation. The DI estimator drawn from our framework
is fully efficient with the firm theoretical ground. Moreover, the
weighted-bootstrap variance estimator is consistent with straightforward
realization and the avoidance of re-imputing the missing components
compared to the conventional bootstrap methods. In
the motivating example in Section \ref{sec:example}, DI resolves
the overestimation issue of Rubin's combining rule under MI in the
sensitivity analysis and detects a statistically significant benefit
of using the test drug to cure depression. Our framework is applicable
to a wide range of sensitivity models defined
through estimating equations.

The rest of the paper proceeds as follows. Section
\ref{sec:example} uses antidepressant clinical trial data to motivate
the development of an efficient imputation method. Section \ref{sec:setup}
introduces the basic setup, provides notations, estimands, imputation
mechanisms in sensitivity analyses, and comments on existing methods
to handle missingness. Section \ref{sec:fi} presents DI and its main
steps. Section \ref{sec:theory} gives the asymptotic theories for
the DI estimator and proposes weighted bootstrap on variance estimation.
Section \ref{sec:simulation} explores the finite-sample performance
of the DI estimator via simulation. Section \ref{sec:application}
returns to the motivating example and applies the proposed framework
to the data. Section \ref{sec:conclude} draws the conclusion. Supplementary
material contains the technical setup, proof of the theorems, and
additional simulation and real-data application results.

\section{Motivating example \label{sec:example}}

An antidepressant clinical trial from the Auspices
of the Drug Information Association is conducted to evaluate the effect
of an experimental medication \citep{mallinckrodt2014recent}. The study measures the longitudinal outcomes of the HAMD-17 score at baseline
and weeks 1, 2, 4, 6, and 8 for 200 patients who are randomly assigned to the control and treatment groups at a 1:1 ratio. The data has a monotone missingness pattern except for one individual in the treatment group containing intermittent missingness. For illustration purposes, we assume MAR for this intermittent missingness and focus on the monotone missing data. 
% The study measures the longitudinal outcomes of the HAMD-17 score at baseline and weeks 1, 2, 4, 6, and 8 for 200 patients who are randomly assigned to the control and treatment groups at a 1:1 ratio. Within the study, all missingness happens in a monotone pattern except one individual in the treatment group containing intermittent missingness. For illustration purposes, we assume MAR for this intermittent missingness and focus on the monotone missing data. 
To investigate the treatment effect
in different aspects, we explore two population summaries by constructing different treatment effect estimands to evaluate the treatment effect. The first population-level summary is
the average treatment effect (ATE) defined by the difference between the relative change of the
HAMD-17 score from the baseline value in the last visit. The second population-level summary is the risk difference defined
by the percentage difference of patients with $50\%$ or more improvement
from the baseline HAMD-17 score at the last visit. % We define the ATE and risk difference formally  in Section \ref{sec:setup}. 

\begin{figure}
\caption{Spaghetti plots of the relative change of the HAMD-17 score separated
by the two treatments. \label{fig:sp_real}}

\centering{}\includegraphics[scale=0.35]{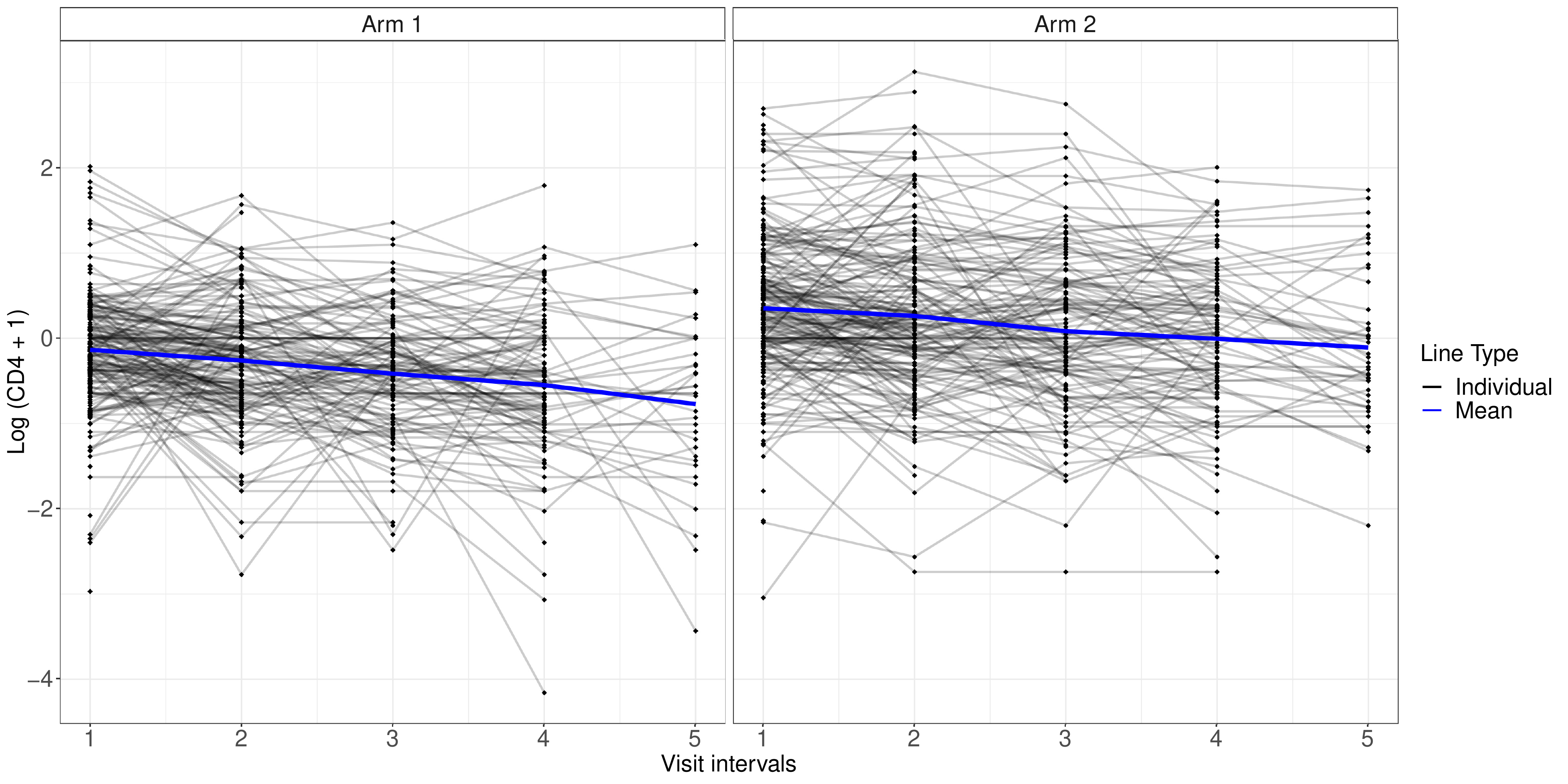}
\end{figure}

Figure \ref{fig:sp_real} shows the spaghetti plot of
the relative change of the HAMD-17 score for the two groups. It reflects
a typical missing data issue in longitudinal clinical trials, with
39 patients in the control group and 30 patients in the treatment
group dropping out during the study period. The relatively high dropout
rate prompts the need for imputation to utilize the information related
to the missingness. 

\begin{figure}[!htbp]
\caption{Normal Q-Q plots of the conditional residuals at each visit. \label{fig:diagnosis}}

\centering{}\includegraphics[scale=0.5]{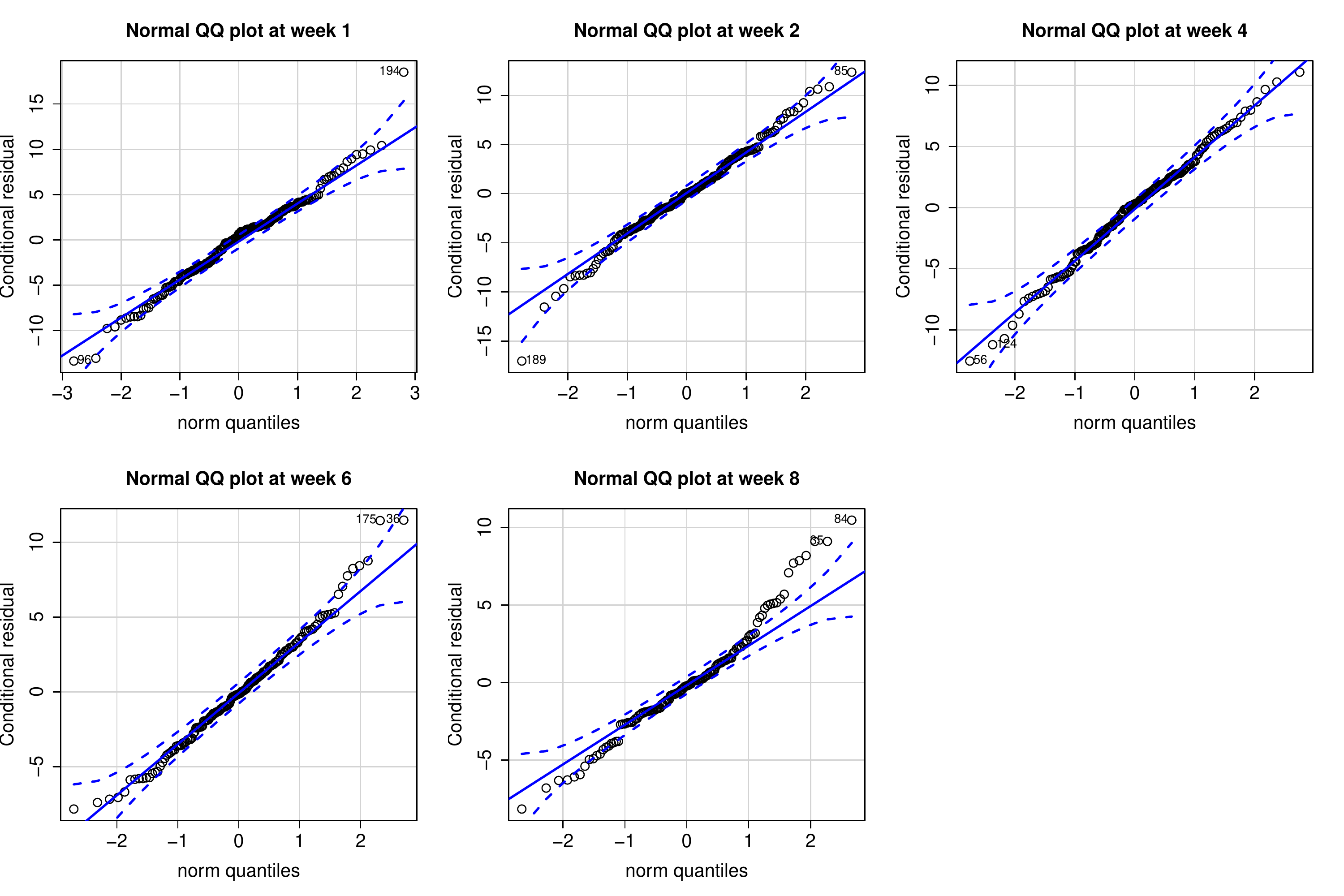} 
\end{figure}

As MI relies on the parametric modeling assumption,
we begin by checking the normality of the conditional errors at each
visit for model diagnosis. Figure \ref{fig:diagnosis} illustrates
the normal Q-Q plots of the conditional residuals obtained by fitting
the MMRM for the observed data. All the plots indicate an underlying
normal distribution for the outcomes since the majority of residuals
is within the confidence region, only the conditional errors at week
8 being slightly right-skewed.

\begin{figure}[!htbp]
\caption{Estimation results of the ATE and the risk difference under MAR and
J2R, accompanied with the $95\%$ CIs. \label{fig:ci}}

\centering{}\includegraphics[scale=0.5]{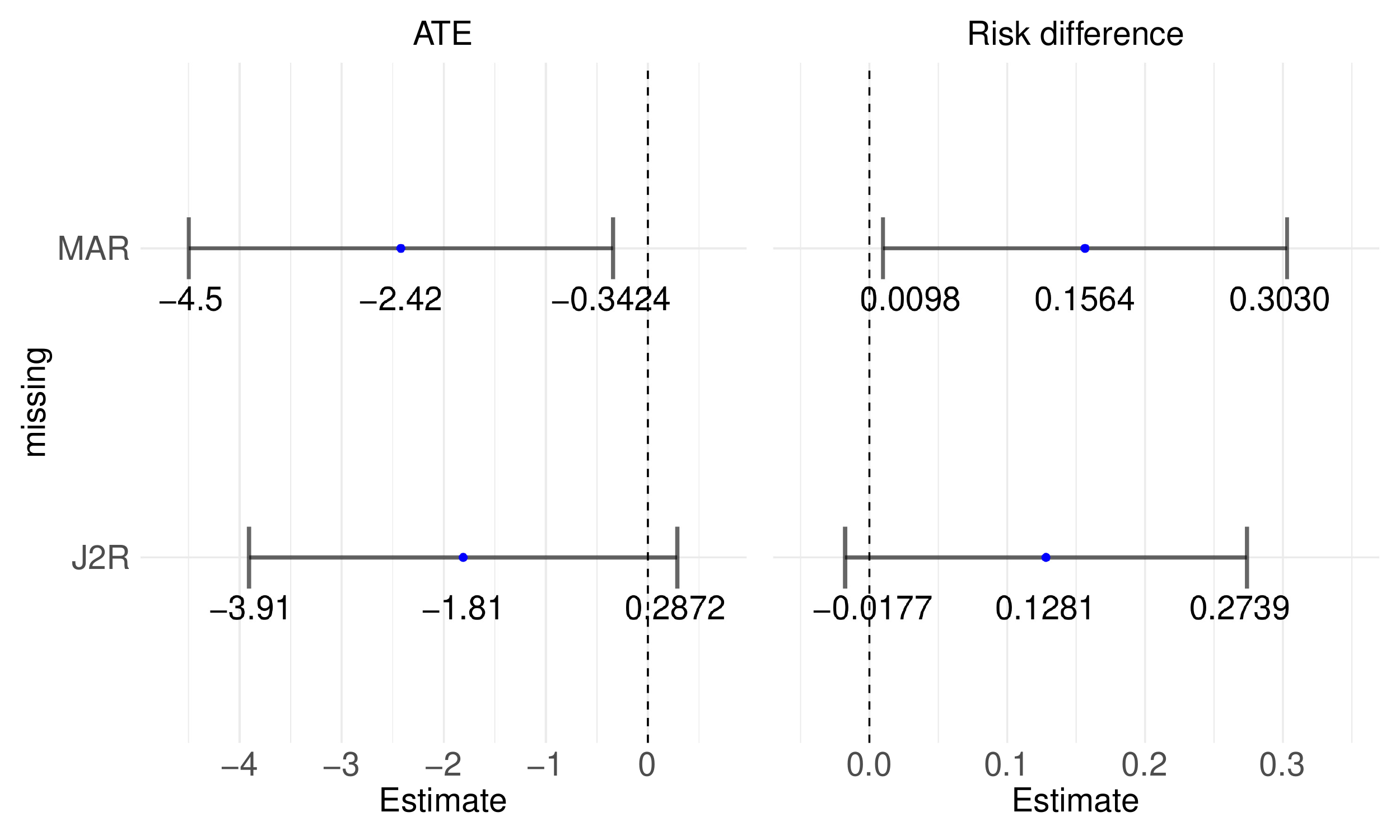}
\end{figure}

Under the normal assumption, we conduct
MI to the missing components with the imputation size as $100$ under
MAR to perform the primary analysis and under J2R for the sensitivity
analysis to obtain the inferences using Rubin's rule.  The point estimates
of the ATE and the risk difference accompanied with the $95\%$ CIs
are presented in Figure \ref{fig:ci}. In the primary analysis that
assumes MAR, both the ATE and the risk difference reveal a significant
treatment effect.  However, the sensitivity analysis under J2R fails
to capture the same significance under MI, leading to a loss of the
credibility of the experimental drug and a potential influence on
the decision made by the investigators. The alteration of the study
conclusion may result from the overestimation issue of Rubin's MI
variance estimator as detected in the literature involving sensitivity
analyses (e.g., \citealp{liu2016analysis}), rather than the loss
of effectiveness in the test drug. To further explore the cause of
the altered study result, it is vital to overcome the overestimation
issue brought up from MI and develop an efficient imputation approach
to obtain a consistent variance estimator without an expensive computational
cost.

\section{Basic setup \label{sec:setup}}

\subsection{Notations and estimands \label{sec:notation}}

Let $Y_{ik}$ be the continuous response for patient $i$ at time
$t_{k}$, where $i=1,\cdots,n$, and $k=1,\cdots,T$. Denote the baseline
$p$-dimensional fully-observed covariate vector as $X_{i}$, the
group indicator as $G_{i}$ ranging from 1 to $J$ to represent $J$
distinct treatment groups, and the observed indicator as $R_{i}=(R_{i1},\cdots,R_{iT})^{\intercal}$
for patient $i$, where $R_{ik}=1$ if $Y_{ik}$ is observed, $R_{ik}=0$
otherwise. Without loss of generality, we consider $J=2$, where $G_{i}=1,2$
represents the $i$th patient is in the control or active treatment
group, respectively. Let $Y_{i}=(Y_{i1},\cdots,Y_{iT})^{\intercal}$
be a $T$-dimensional longitudinal vector containing history and current
information. A monotone missingness pattern is assumed, i.e., if missingness
begins at time $t^{*}$, we have $R_{ik}=1$ for $t_{k}<t^{*}$ and
$R_{ik}=0$ for $t_{k}\geq t^{*}$. We can partition each response
as $Y_{i}=(Y_{\text{obs},i},Y_{\text{mis},i})^{\intercal}$, where
$Y_{\text{obs},i}$ and $Y_{\text{mis},i}$ are the observed and missing
components. Denote $Z_{i}=(Y_{i},X_{i},G_{i},R_{i})$ as the full
data for patient $i$. In the presence of missing data, denote $Z_{\text{obs},i}=(Y_{\text{obs},i},X_{i},G_{i},R_{i})$
as the observed data. Then, $Z_{i}=(Z_{\text{obs},i},Z_{\text{mis},i})$
corresponds to the combination of the observed and missing parts.

For the treatment comparison, we consider different treatment effect
estimands based on different population-level summaries defined through the estimation equations as $\tau=\tau_{2}-\tau_{1}$,
where $\tau_{j}$ is the value such that $\E\left\{ \psi_{j}(Z_{i},\tau_{j})\right\} =0$
and $\psi_{j}(Z,\tau_{j})$ is a function in the space $\mathbb{R}^{d}\times\Omega$,
with $\Omega$ as a compact subset of a Euclidean space and $\psi_{j}(Z,\tau_{j})$
as a continuous function of $\tau_{j}$ for each vector $Z$ and measurable
of $Z$ for each $\tau_{j}$. 

The expectation $\E\left\{ \psi_{j}(Z_{i},\tau_{j})\right\} $
prompts the determination of the full-data distribution. Under MNAR,
we use the PMM framework to describe the data distribution as $f(Z)=\int f(Z\mid R)f(R)dR$,
where $Z_{1},\cdots,Z_{n}$ are i.i.d. from a parametric model $f(Z,\theta_{0})$
with the support free of $\theta_{0}$. Here, $Z\in\mathbb{R}^{d}$
is a $d$-dimensional vector, $\Theta$ is a $q$-dimensional Euclidean
space, and $\theta_{0}$ is the true model parameter lying in the
interior of $\Theta$. Moreover, let $s(Z,\theta)$ be the score function
of $f(Z,\theta)$, and assume $s(Z,\theta)$ to be a continuous function
of $\theta$ for each $Z$ and a measurable function of $Z$ for each
$\theta$. The identification of the treatment effect relies on the
assumption of the pattern-specific data distribution $f(Z\mid R)$
under each missingness pattern, which is prespecified in a statistical analysis plan for a
clinical trial as the sensitivity models under hypothetical scenarios
to address potential intercurrent events that may impact the estimation of treatment effect \citep{international2019addendum}.

The most popular full-data model in longitudinal
clinical trials is the MMRM as recommended by the FDA and National
Research Council \citep{little2012prevention}, which assumes an underlying
multivariate normal distribution for the longitudinal outcomes. The
motivating example in Section \ref{sec:example} validates its application
in practice. \textcolor{black}{Therefore, throughout the paper, we
assume that the continuous longitudinal outcomes $Y_{i}$ given the
covariates $X_{i}$ and the group indicator $G_{i}=j$ independently
follow a multivariate normal distribution as 
\begin{equation}
Y_{i}\mid(X_{i},G_{i}=j)\sim\mathcal{N}_{T}(\mu_{ij},\Sigma^{(j)}),\label{eq:general_model}
\end{equation}
where $\mu_{ij}=(\mu_{ij1},\cdots,\mu_{ijT})^{\intercal}=(\Tilde{X}_{i}^{\intercal}\beta_{j1},\cdots,\Tilde{X}_{i}^{\intercal}\beta_{jT})^{\intercal}$,
$\beta_{j1},\cdots,\beta_{jT}$ are $(p+1)$ dimensional group-specific
vectors, $\Tilde{X}_{i}=(1,X_{i}^{\intercal})^{\intercal}$, and $\Sigma^{(j)}$
is a group-specific covariance matrix. }

When no missingness is involved in the data, the
only pattern corresponds to $R=\mathbf{1}_{T}$, where $\mathbf{1}_{T}$
is a $T$-dimensional all-ones vector representing the outcomes are
fully observed. The treatment effect identification boils down to
the specification of the conditional distribution $Y_{i}\mid(X_{i},G_{i}=j)$,
which has been specified in the formula \eqref{eq:general_model}.
Under this circumstance, we present three typical population-level summaries to define the treatment effect estimands in the following example. 
% For simplicity, we refer to them as the treatment effect estimands in this paper.

\begin{example}[Treatment effect estimands]\label{example1} The
parameter $\tau=\tau_{2}-\tau_{1}$ can represent different types
of treatment effect given the following choices of $\psi_{j}(\cdot)$
for $j=1,2$: 
\begin{enumerate}
\item The ATE when $\tau_{j}=\mathbb{E}(Y_{iT}\mid G_{i}=j)$ for $j=1,2$:
$\psi_{j}(Z_{i},\tau_{j})=I(G_{i}=j)(Y_{iT}-\tau_{j})$. 
\item The risk difference when $\tau_{j}=P(Y_{iT}\geq c\mid G_{i}=j)$ for
$j=1,2$: $\psi_{j}(Z_{i},\tau_{j})=I(G_{i}=j)\big\{ I(Y_{iT}\geq c)-\tau_{j}\big\}$,
where $c$ is a prespecified threshold. 
\item Distributional information of the treatment, e.g., the QTE for the
$q$th quantile of responses when $\tau_{j,q}$ is the $q$th quantile
for distribution of outcomes at the last time point for $j=1,2$:
$\psi_{j}(Z_{i},\tau_{j,q})=I(G_{i}=j)\big\{ I(Y_{iT}\leq\tau_{j,q})-q\big\}$. 
\end{enumerate}
\end{example}

Among the above treatment effect estimands, the ATE is most widely
used to describe the treatment effect in clinical trials (e.g., \citealp{carpenter2013analysis};
\citealp{liu2016analysis}; \citealp{zhang2020likelihood}). Some
clinical studies also care about the risk difference regarding the
percentage of patients with the endpoint continuous outcomes dichotomized
by a certain threshold for each group. For example, \citet{roussel2019double}
take the difference of the percentage of participants with HbA1c $<7\%$
in each group as a secondary endpoint. However, the ATE is possibly
insufficient to capture the effect of treatment under a skewed outcome
distribution, where treatment may not influence the average outcome,
but the tail of the outcome distribution. In these cases, the QTE
is preferred \citep{yang2020multiply}. 

\subsection{Sensitivity models in sensitivity analyses \label{sec:imp_mechanism}}

When there is more than one missingness pattern,
the identification of the treatment effect is accomplished by specifying
the data distribution $f(Z\mid R)$ under each pattern. Since the
distribution $f(Z\mid R)$ is unobserved if $R\neq\mathbf{1}_{T}$,
several plausible sensitivity models are proposed to model the missingness.
\citet{international2019addendum} addresses the importance of specifying explicit MNAR assumptions that
underlie the sensitivity analysis in advance of the clinical trial
based on different characteristics of drugs.
Here, we concentrate on the J2R assumption to assess the robustness
of the study conclusions.

The J2R sensitivity model is one specific control-based
imputation model, which envisions the missing responses in the treatment
group will have the same outcome profile as those in the control group
with the same covariates after dropout \citep{carpenter2013analysis}.
 It can be a conservative missingness assumption 
that assumes the average treatment effect disappears immediately after patients discontinue from active treatment, and it is a commonly used sensitivity analysis in practice for longitudinal clinical trials (\citealp{mallinckrodt2014recent}; \citealp{mallinckrodt2019estimands}).

Equipped with the normality assumption, the group-specific model for
the missing outcomes is the conditional distribution given the observed
data under each missingness pattern as
\begin{align}
Y_{\text{mis},i} & \mid(Y_{\text{obs},i},X_{i},G_{i}=j,R_{ik-1}=1,R_{ik}=0)\sim\nonumber \\
 & \mathcal{N}_{T-k}\big\{\mu_{\text{mis},ij}^{(k)}+\Sigma_{21}^{(1)}\Sigma_{11}^{(1)-1}(Y_{\text{obs},i}-\mu_{\text{obs},ij}^{(k)}),\Sigma_{22}^{(1)}-\Sigma_{21}^{(1)}\Sigma_{11}^{(1)-1}\Sigma_{12}^{(1)}\big\},\label{eq:j2r_model}
\end{align}
where $\mu_{\text{mis},ij}^{(k)},\mu_{\text{obs},ij}^{(k)}$ are the
individual-specific mean vectors and $\Sigma^{(1)}=\begin{pmatrix}\Sigma_{11}^{(1)} & \Sigma_{12}^{(1)}\\
\Sigma_{21}^{(1)} & \Sigma_{22}^{(1)}
\end{pmatrix}$ is the covariance matrix for the control group partitioned corresponding
to $Y_{\text{obs},i}$ and $Y_{\text{mis},i}$. Therefore, modeling the J2R sensitivity model corresponds to specifying the group-specific mean vector and covariance.

\begin{assumption}[Jump-to-reference model]\label{assump:j2r}
For the control group, the missing components are MAR. The imputation
model is of the form \eqref{eq:j2r_model}, with $\mu_{\text{mis},i1}^{(k)}=(\Tilde{X}_{i}^{\intercal}\beta_{1(k+1)},\cdots,\allowbreak \Tilde{X}_{i}^{\intercal}\beta_{1T})^{\intercal}$
and $\mu_{\text{obs},i1}^{(k)}=(\Tilde{X}_{i}^{\intercal}\beta_{11},\cdots,\Tilde{X}_{i}^{\intercal}\beta_{1k})^{\intercal}$.

For the treatment group, the imputation model is of the form \eqref{eq:j2r_model},
with $\mu_{\text{mis},i2}^{(k)}=(\Tilde{X}_{i}^{\intercal}\beta_{1(k+1)},\cdots,\allowbreak \Tilde{X}_{i}^{\intercal}\beta_{1T})^{\intercal}$
and $\mu_{\text{obs},i2}^{(k)}=(\Tilde{X}_{i}^{\intercal}\beta_{21},\cdots,\Tilde{X}_{i}^{\intercal}\beta_{2k})^{\intercal}$
representing the regression coefficients for the participants after
deviation will "jump to" the ones in the control group with the
same baseline covariates. \end{assumption}

Apart from J2R as a way to quantify the deviation from MAR, we also
summarize two more MNAR assumptions as the RTB and washout imputation
used in FDA statistical review and evaluations reports (e.g., \citealp{rtb2016tresiba}) to represent
different ways to model the missing outcomes regardless of compliance
after discontinuation in sensitivity analyses. The RTB approach assumes a washout effect for the missing responses at
the last time point in both groups, indicating that the outcome of interest will return
to the baseline performance regardless of the prior treatment after
dropout. However, the biological plausibility of the washout assumption needs to be carefully evaluated, and RTB is not necessarily conservative when missing data is imbalanced between treatment and control group \citep{zhang2020missing}.

\begin{assumption}[Return-to-baseline model]\label{assump:rtb}
 The imputation model of the outcomes at the last time point follows
the marginal baseline model $Y_{iT}\mid(X_{i},R_{iT}=0,G_{i}=j)\sim\mathcal{N}(\mu_{ij1},\Sigma_{(1,1)}^{(j)})$,
where $\Sigma_{(1,1)}^{(j)}$ represents the $(1,1)$ element of $\Sigma^{(j)}$.
\end{assumption}

Washout imputation also acts as a possible sensitivity
model and appears in several statistical review and evaluation reports
(e.g., \citealp{washout2018praluent}). It combines
the idea of the RTB and J2R assumptions by assuming a MAR pattern for
the control group and an RTB pattern for the missing outcomes in the treatment group. 
One of the reasons to consider washout imputation is to address the potential issue of imbalanced missing data in RTB.

\begin{assumption}[Washout model]\label{assump:washout} For the
control group, the assumption for the missing responses is MAR. The
model is the same as the one for control group in Assumption \ref{assump:j2r}.
For the treatment group, the assumption for the missing responses
is the same as Assumption \ref{assump:rtb}. \end{assumption}

Given a prespecified sensitivity model that characterizes
the MNAR assumption, one can therefore determine the pattern-specific
data distribution $f(Z\mid R)$ and identify the treatment effect
under the PMM framework. \textcolor{black}{To obtain valid treatment
effect inferences, one can }implement the conventional likelihood-based
or imputation approach to deal with missingness in sensitivity analyses.
Both methods are elaborated on next section.

\subsection{Existing methods to handle missingness in sensitivity analyses}

The likelihood-based method and MI are two traditional approaches
to handle missingness in sensitivity analyses in longitudinal clinical
trials. The likelihood-based method utilizes the
MMRM model and the ignorability of the missing components under MAR
to draw valid inferences. In terms of the MNAR-related sensitivity
models, the analytical form of the inference is obtained via averaging
over the dropout patterns based on the PMM framework (e.g., \citealp{liu2016analysis,mehrotra2017missing}).
However, the treatment effect estimator can be infeasible to derive
under cases where the normality assumption is violated or the estimand
of interest is not of mean type. For example, when we focus on the
risk difference of the test drug in the antidepressant trial in Section
\ref{sec:example}, complexity arises when incorporating the dropout
patterns. Moreover, the likelihood-based estimator needs to be re-derived
for different imputation mechanisms and can be complicated if there
are multiple missingness patterns.

MI provides a simple way to handle diverse types of estimands. It
creates multiple complete datasets by conducting imputations based
on the prespecified imputation model and has Rubin's combining rule
to obtain inferences. We illustrate one typical strategy
for conducting MI with Rubin's rule, which has appeared in the literature
(e.g., \citealp{carpenter2013analysis,mallinckrodt2020aligning}),
in the sensitivity analysis under J2R in longitudinal clinical trials
using the estimands in Example \ref{example1} as follows. 
\begin{description}
\item [{{Step 1}.}] For the observed
data in the control group, fit the MMRM and obtain the estimated sensitivity
model.
\item [{{Step 2}.}] Impute the missing
values in both groups from the sensitivity model specified in Assumption
\ref{assump:j2r}. Repeat the imputation $M$ times to create $M$ imputed datasets. 
\item [{{Step 3}.}] For each imputed
dataset, conduct a complete data analysis by solving the estimating
equations that correspond to the estimands in Example \ref{example1}
to obtain $\hat{\tau}_{\text{MI}}^{(m)}$ as the estimator of the
$m$th imputed dataset, where $m=1,\cdots,M$. 
\item [{{Step 4}.}] Combine the estimations from the $M$ imputed datasets
by Rubin's combining rule and obtain the MI estimator as $\hat{\tau}_{\text{MI}}=M^{-1}\sum_{m=1}^{M}\hat{\tau}_{\text{MI}}^{(m)}$,
with the variance estimator 
\[
\mathbb{\hat{V}}(\hat{\tau}_{\MI})=\frac{1}{M}\sum_{m=1}^{M}\mathbb{\hat{V}}(\hat{\tau}_{\text{MI}}^{(m)})+\left(1+\frac{1}{M}\right)B_{M},
\]
where $B_{M}=(M-1)^{-1}\sum_{m=1}^{M}(\hat{\tau}_{\MI}^{(m)}-\hat{\tau}_{\MI})^{2}$
represents the between-imputation variance. 
\end{description}
\citet{wang1998large} discover inefficiency in point estimation in
the general MI procedure. More loss of efficiency occurs in interval
estimation, where Rubin's method may overestimate the variance in
practice \citep{robins2000inference}. To illustrate the problem,
the variance of the MI estimator is 
\[
\mathbb{V}(\hat{\tau}_{\MI})=\mathbb{V}(\hat{\tau}_{\MI}-\hat{\tau})+\mathbb{V}(\hat{\tau})+2\text{cov}(\hat{\tau}_{\MI}-\hat{\tau},\hat{\tau}),
\]
where $\hat{\tau}$ is the treatment effect estimator for the fully
observed data. Rubin's method only estimates $\mathbb{V}(\hat{\tau}_{\MI}-\hat{\tau})$
and $\mathbb{V}(\hat{\tau})$, and it treats $\text{cov}(\hat{\tau}_{\MI}-\hat{\tau},\hat{\tau})=0$
which does not hold in sensitivity analyses that assume MNAR. For
example, \citet{liu2016analysis} find that the variance estimator
using Rubin's rule tends to overestimate the true variance in simulation
studies under J2R in sensitivity analyses. The motivating
example in Section \ref{sec:example} further captures a change in
the study result using MI with Rubin's rule under J2R in the sensitivity
analysis, which may result from the overestimation issue. One approach
to deal with this issue is to use bootstrap to derive the replication-based
variance estimation. But it is computationally intensive since the
missing values have to be re-imputed $M$ times based on the reconstructed
sensitivity model in each bootstrap step.

Therefore, a more efficient method to get valid estimators for diverse
kinds of treatment effect estimands and the corresponding appropriate
variance estimators with a simple implementation is needed. We propose
DI based on the idea of MC integration to get the inference and the
weighted bootstrap procedure to obtain a consistent variance estimator. 

\section{Distributional imputation \label{sec:fi}}

We propose DI to draw the inference on the treatment effect in sensitivity
analyses. Given the parametric distributions of the missing components
based on certain sensitivity analysis settings, the key insight is
to impute each missing value by samples from its conditional distribution
given the observed data. Drawn on the idea of MC integration, any
estimating equations applied to the imputed dataset approximate the
mean estimating equations given the observed data and thus allow an
efficient estimation of the target estimand.

The use of the mean estimating equations conditional
on the observed data to assess the treatment effect is prevalent in
the missing data literature. \citet{louis1982finding} takes advantage
of the conditional mean estimating equations with the expectation-maximization
algorithm to obtain valid inferences for the incomplete data. \citet{robins2000inference}
also apply the idea of mean estimating equations to allow for the
incompatibility between the imputation and analysis model. In the
presence of missing data, one can estimate the function $\psi_{j}(Z_{i},\tau_{j})$
that characterizes the treatment effect by the conditional expectation
given the observed values under certain sensitivity models that have
been prespecified in the trial protocol. Therefore, a consistent
estimator of $\tau_{j}$ for $j=1,2$ is the solution to 
\begin{equation}
\sum_{i=1}^{n}\mathbb{E}\{\psi_{j}(Z_{i},\tau_{j})\mid Z_{\text{obs},i},\hat{\theta}\}=0,\label{conditional_exp}
\end{equation}
where $\hat{\theta}$ is a consistent estimator of an unknown modeling
parameter $\theta\in\Theta$. A common choice of $\hat{\theta}$ is
the pseudo maximum likelihood estimator (MLE) given the observed data,
i.e., it solves the mean score equations
\begin{equation}
\frac{1}{n}\sum_{i=1}^{n}\mathbb{E}\{s(Z_{i},\theta)\mid Z_{\text{obs},i}\}=0.\label{theta_score_eq}
\end{equation}
Note that the mean estimating equations in \eqref{conditional_exp}
and \eqref{theta_score_eq} have general forms which can accommodate
different sensitivity models. In longitudinal clinical trials, the
commonly used mean estimating equations correspond to the score function
of the MMRM for the observed data. However, even under the multivariate
normal assumption, the explicit form of the estimator $\hat{\tau}_{j}$
is only feasible to obtain when the function $\psi_{j}(Z_{i},\tau_{j})$
has a linear form such as the one in Example \ref{example1} (a).

We can estimate the conditional expectation $\mathbb{E}\{\psi_{j}(Z_{i},\tau_{j})\mid Z_{\text{obs},i},\hat{\theta}\}$
using the complete data after imputation. For the missing component
of $i$th continuous response $Y_{i}$, we independently draw $Y_{\text{mis},i}^{*(1)},\cdots,Y_{\text{mis},i}^{*(M)}$
from a prespecified sensitivity model with the estimated conditional
distribution $f(Z_{\text{mis},i}^{*(m)}\mid Z_{\text{obs},i},\hat{\theta})$
such as Assumptions \ref{assump:j2r}--\ref{assump:washout} used
in sensitivity analyses.

With the imputed data, denote $Y_{i}^{*(m)}=(Y_{\text{obs},i},Y_{\text{mis},i}^{*(m)})$
as the imputed longitudinal responses and $Z_{i}^{*(m)}=(Z_{\text{obs},i},Z_{\text{mis},i}^{*(m)})$
as the full imputed data for $i$th patient. DI incorporates the idea
of MC integration. When the imputation size $M$ is large, one can
estimate the conditional expectation in \eqref{conditional_exp} as
\begin{equation}
\mathbb{E}\{\psi_{j}(Z_{i},\tau_{j})\mid Z_{\text{obs},i},\hat{\theta}\}\approx\frac{1}{M}\sum_{m=1}^{M}\psi_{j}(Z_{i}^{*(m)},\tau_{j}).\label{weighted_conditional}
\end{equation}

Based on the MC approximation, we can therefore derive the DI estimator
$\hat{\tau}_{\DI,j}$ for $j$th group by solving the estimating equations
\begin{equation}
\frac{1}{M}\sum_{i=1}^{n}\sum_{m=1}^{M}\psi_{j}(Z_{i}^{*(m)},\tau_{j})=0.\label{fi_ee}
\end{equation}

\begin{example}[DI estimator for the treatment effect]\label{example2}
For all estimands in Example \ref{example1}, the DI estimator of
the treatment effect is $\hat{\tau}_{\DI}=\hat{\tau}_{\DI,2}-\hat{\tau}_{\DI,1}$,
where $\hat{\tau}_{\DI,j}$ for $j=1,2$ is derived by defining the
following specific $\psi_{j}$ function and solving the following
estimating equations: 
\begin{enumerate}
\item The ATE: Set $\psi_{j}(Z_{i}^{*(m)},\tau_{j})=I(G_{i}=j)\big\{ R_{iT}Y_{\text{obs},iT}+(1-R_{iT})Y_{\text{mis},iT}^{*(m)}-\tau_{j}\big\}$,
and $\hat{\tau}_{\DI,j}$ is the solution to 
\[
\sum_{i=1}^{n}I(G_{i}=j)\left\{ R_{iT}Y_{\text{obs},iT}+(1-R_{iT})(M^{-1}\sum_{m=1}^{M}Y_{\text{mis},iT}^{*(m)})-\tau_{j}\right\} =0.
\]
\item The risk difference of the treatment: Set 
\[
\psi_{j}(Z_{i}^{*(m)},\tau_{j})=I(G_{i}=j)\big\{ R_{iT}I(Y_{\text{obs},iT}\geq c)+(1-R_{iT})I(Y_{\text{mis},iT}^{*(m)}\geq c)-\tau_{j}\big\},
\]
and $\hat{\tau}_{\DI,j}$ is the solution to 
\[
\sum_{i=1}^{n}I(G_{i}=j)\Big[R_{iT}I(Y_{\text{obs},iT}\geq c)+(1-R_{iT})\left\{ M^{-1}\sum_{m=1}^{M}I(Y_{\text{mis},iT}^{*(m)}\geq c)\right\} -\tau_{j}\Big]=0.
\]
\item Distributional information of the treatment, e.g. the QTE for the
$q$th quantile of responses when $\tau_{j,q}$ is the $q$th quantile
for distribution of outcomes at the last time point: Set
\[
\psi_{j}(Z_{i}^{*(m)},\tau_{j,q})=I(G_{i}=j)\big\{ R_{iT}I(Y_{\text{obs},iT}\leq\tau_{j,q})+(1-R_{iT})I(Y_{\text{mis},iT}^{*(m)}\leq\tau_{j,q})-q\big\},
\]
and $\hat{\tau}_{\DI,j}$ is the solution to 
\[
\sum_{i=1}^{n}I(G_{i}=j)\Big[R_{iT}I(Y_{\text{obs},iT}\leq\tau_{j,q})+(1-R_{iT})\left\{ M^{-1}\sum_{m=1}^{M}I(Y_{\text{mis},iT}^{*(m)}\leq\tau_{j,q})\right\} -q\Big]=0.
\]
\end{enumerate}
\end{example}

\begin{remark}[Discussion of the incorporation of covariates in
estimation] In sensitivity analyses, one may incorporate the covariate
information to improve the efficiency of the treatment effect estimator
\citep{tsiatis2007semiparametric}. For example, the ATE estimator
derived from the sample average may not be the most efficient one;
the one motivated by the analysis of covariance model (ANCOVA) is
preferred in practice. We present an ANCOVA-motived DI estimator $\hat{\tau}_{\DI,j}$
by defining the $\psi_{j}$ function and its corresponding estimating
equations for $j=1,2$ as follows:

Set the $\psi_{j}$ function as 
\[
\psi_{j}(Z_{i}^{*(m)},\tau_{j},\gamma)=\begin{pmatrix}V_{i}\big\{ R_{iT}Y_{\text{obs},iT}+(1-R_{iT})Y_{\text{mis},iT}^{*(m)}-V_{i}^{\intercal}\gamma\big\}\\
\begin{Bmatrix}\Tilde{X}_{i}^{\intercal} & I(j=2)\Tilde{X}_{i}^{\intercal}\end{Bmatrix}\gamma-\tau_{j}
\end{pmatrix},
\]
where $V_{i}=\begin{Bmatrix}\Tilde{X}_{i}^{\intercal} & I(G_{i}=2)\Tilde{X}_{i}^{\intercal}\end{Bmatrix}^{\intercal}$,
and $\gamma$ is the vector of joint regression coefficients in the
two treatment groups. $\hat{\tau}_{\DI,j}$ is the solution to 
\[
\sum_{i=1}^{n}\begin{pmatrix}V_{i}\big\{ R_{iT}Y_{\text{obs},iT}+(1-R_{iT})(M^{-1}\sum_{m=1}^{M}Y_{\text{mis},iT}^{*(m)})-V_{i}^{\intercal}\gamma\big\}\\
\begin{Bmatrix}\Tilde{X}_{i}^{\intercal} & I(j=2)\Tilde{X}_{i}^{\intercal}\end{Bmatrix}\gamma-\tau_{j}
\end{pmatrix}=0.
\]

Note that the ANCOVA-motivated estimator for the ATE replaces the
treatment-specific covariate mean with the overall covariate mean
to gain efficiency, while this information is not applicable to the
risk difference and the QTE. Without this information, the estimating
functions presented in Example \ref{example2} render the most efficient
estimators of the ATE, the risk difference and the QTE. One can use
the augmented inverse propensity weighted (AIPW) type of estimators
to incorporate additional information in the propensity score and
outcome regression model \citep{zhang2012causal}; however, AIPW does
not improve the efficiency of the simple estimators (supported by
unshown simulation studies). \end{remark}

To summarize, the DI procedure under specific sensitivity models is as follows: 
\begin{description}
\item [{{Step 1}.}] For each group, fit an MMRM from a population-averaged
perspective for the observed data and obtain the model parameter estimator
$\hat{\theta}$ by solving the estimating equations \eqref{theta_score_eq}. 
\item [{{Step 2}.}] Impute the missing values $M$ times from the estimated
imputation model $f(Z_{\text{mis},i}^{*(m)}\mid Z_{\text{obs},i},\hat{\theta})$
based on prespecified imputation mechanisms such as Assumptions \ref{assump:j2r}--\ref{assump:washout}
for each group. 
\item [{{Step 3}.}] Derive the DI estimator $\hat{\tau}_{\DI,j}$ by solving
the estimating equations \eqref{fi_ee} and get the
treatment effect DI estimator $\hat{\tau}_{\DI}$. 
\end{description}
The theoretical asymptotic properties of the DI estimator and the
variance estimation procedure are given in Section \ref{sec:theory}.

\begin{remark}[Computation complexity of DI and MI] DI and MI both
use $M^{-1}$ as the weight for each imputed dataset. However, the
approaches to conduct the full-data analysis after the imputation
procedure are different. For MI, we conduct separate analyses for
each imputed dataset and use Rubin's MI combining rule to get inference;
while for DI, the analysis is done jointly based on the entire imputed
dataset, with the inference derived from the mean estimating equations
\eqref{fi_ee}. In terms of the computation complexity after imputation,
DI is more computationally efficient than MI for point estimation.
For example, if linear models are fitted to in the analysis stage,
MI fits $M$ linear models separately, with the total computational
cost $O(Mnp^{2}+Mp^{3})$; DI fits one linear model for the pooled
imputed dataset, with the total computational cost $O(Mnp^{2}+p^{3})$
\citep{friedman2001elements}. \end{remark}

\begin{remark}[Connection with fractional imputation] DI is similar
to parametric fractional imputation (FI; \citealp{kim2011parametric};
\citealp{yang2016fractional}), where we pool the $M$ imputed dataset
and conduct the full-data analysis jointly by solving the estimating
equations. Our proposed DI utilizes the distributional behavior of
the missing components, by imputing them directly from the estimated
conditional distribution given the observed data under some prespecified
sensitivity analysis settings, thus avoids applying importance sampling
required by FI to generate imputed data from the proposal distribution,
and yields simplicity. \end{remark}

\begin{remark}[Choice of the imputation size $M$] DI utilizes
the idea of MC integration to approximate the conditional expectation
in \eqref{conditional_exp}. Based on the MC approximation theory
\citep{geweke1989bayesian}, the MC error rate is $O(M^{-1/2})$ for
any dimension. If the model is not computationally intensive, larger
$M$ can be selected to further reduce the MC error. As shown in the
simulation studies, the selection of the imputation size $M$ is not
sensitive to the inferences. We observe a decent performance of the DI
estimator with a small imputation size $M$ (e.g., $M=5$). \end{remark}

\section{Theoretical properties and variance estimation \label{sec:theory}}

\subsection{Asymptotic properties of the DI estimator}

We verify the consistency and asymptotic normality for the DI estimator.
For simplicity, we consider the inference for one group here and omit
the group subscript $j$. Extension to multiple groups is straightforward.
Denote $\tau_{0}$ as the true parameter such that $\mathbb{E}\big\{\psi(Z_{i},\tau)\big\}=0$.
The comprehensive regularity conditions and technical proofs are given in Sections \ref{appen:thm1} and \ref{appen:thm2}
in the supplementary material.

\begin{theorem}\label{consistency_thm} Under the regularity conditions
listed in Section \ref{appen:thm1} in the supplementary material,
the DI estimator $\hat{\tau}_{\DI}\xrightarrow{\mathbb{P}}\tau_{0}$
as the imputation size $M\rightarrow\infty$ and sample size $n\rightarrow\infty$.
\end{theorem}

\begin{theorem}\label{theo:asym_normal} Under the regularity conditions
listed in Section \ref{appen:thm2} in the supplementary material,
as the imputation size $M\rightarrow\infty$ and sample size $n\rightarrow\infty$,
\[
\sqrt{n}(\hat{\tau}_{\DI}-\tau_{0})\xrightarrow{d}\mathcal{N}[0,A(\tau_{0},\theta_{0})^{-1}B(\tau_{0},\theta_{0})\{A(\tau_{0},\theta_{0})^{-1}\}^{\intercal}],
\]
where 
\begin{align*}
A(\tau_{0},\theta_{0}) & =\mathbb{E}\Big[\frac{\partial\psi(Z_{i},\tau_{0})}{\partial\tau}+\big\{\frac{\partial\Gamma(\tau_{0},\theta_{0})}{\partial\tau}\big\}\bar{s}_{i}(\theta_{0})\Big];\\
B(\tau_{0},\theta_{0}) & =\mathbb{V}\big\{\psi_{i}^{*}(\tau_{0},\theta_{0})+\Gamma(\tau_{0},\theta_{0})^{\intercal}\bar{s}_{i}(\theta_{0})\big\}.
\end{align*}
Here $\psi_{i}^{*}(\tau,\theta)=M^{-1}\sum_{m=1}^{M}\psi(Z_{i}^{*(m)},\tau)$,
$\bar{s}_{i}(\theta)=\mathbb{E}\{s(Z_{i},\theta)\mid Z_{\text{obs},i}\}$,
$\Gamma(\tau,\theta_{0})=\mathbf{I}_{\text{obs}}(\theta_{0})^{-1}\mathbf{I}_{\psi,\text{mis}}(\tau,\theta_{0})$,
where $\mathbf{I}_{\text{obs}}(\theta)=\mathbb{E}\{-\partial\bar{s}_{i}(\theta)/\partial\theta\}$
and $\mathbf{I}_{\psi,\text{mis}}(\tau,\theta)=\mathbb{E}[\{s(Z_{i},\theta)-\bar{s}_{i}(\theta)\}\psi(Z_{i},\tau)]$.
\end{theorem}

\subsection{Variance estimation}

From the result of asymptotic normality of $\hat{\tau}_{\DI}$ in
Theorem \ref{theo:asym_normal}, one consistent variance estimator
of $\hat{\tau}_{\DI}$ under a large imputation size $M$ is 
\[
\hat{\mathbb{V}}_{1}(\hat{\tau}_{\DI})=A_{n}(Z,\hat{\tau}_{\DI},\hat{\theta})^{-1}\big\{\frac{1}{n}B_{n}(Z,\hat{\tau}_{\DI},\hat{\theta})\big\}\big\{ A_{n}(Z,\hat{\tau}_{\DI},\hat{\theta})^{-1}\big\}^{\intercal},
\]
where 
\begin{align*}
A_{n}(Z,\hat{\tau}_{\DI},\hat{\theta}) & =\frac{1}{nM}\sum_{i=1}^{n}\sum_{m=1}^{M}\frac{\partial\psi(Z_{i}^{*(m)},\hat{\tau}_{\DI})}{\partial\tau},\\
B_{n}(Z,\hat{\tau}_{\DI},\hat{\theta}) & =\frac{1}{n}\sum_{i=1}^{n}\big\{\psi_{i}^{*}(\hat{\tau}_{\DI},\hat{\theta})+\hat{\Gamma}(\hat{\tau}_{\DI},\hat{\theta})^{\intercal}\bar{s}_{i}^{*}(\hat{\theta})\big\}\big\{\psi_{i}^{*}(\hat{\tau}_{\DI},\hat{\theta})+\hat{\Gamma}(\hat{\tau}_{\DI},\hat{\theta})^{\intercal}\bar{s}_{i}^{*}(\hat{\theta})\big\}^{\intercal},
\end{align*}
and $\bar{s}_{i}^{*}(\theta)=M^{-1}\sum_{m=1}^{M}s(Z_{i}^{*(m)},\theta)$.
Here 
\[
\hat{\Gamma}(\hat{\tau}_{\DI},\hat{\theta})=\big\{\frac{1}{n}\sum_{i=1}^{n}\bar{s}_{i}^{*}(\hat{\theta})\bar{s}_{i}^{*}(\hat{\theta})^{\intercal}\big\}^{-1}\Big[\frac{1}{nM}\sum_{i=1}^{n}\sum_{m=1}^{M}\big\{ s(Z_{i}^{*(m)},\hat{\theta})-\bar{s}_{i}^{*}(\hat{\theta})\big\}\psi(Z_{i}^{*(m)},\hat{\tau}_{\DI})\Big].
\]

The sandwich form of $\mathbb{\hat{V}}_{1}(\hat{\tau}_{\DI})$ involves
the analytical form of estimated score function $s(Z_{i}^{*(m)},\hat{\theta})$
that is difficult to compute in longitudinal settings. The replication-based
variance estimation is preferred for its simplicity, and it is commonly
obtained by nonparametric bootstrap. But it requires computational
efforts and the re-imputation of the missing components on the refitted
imputation model, especially in a large-scale clinical trial with
numerous participants. We propose weighted bootstrap to obtain a consistent
variance estimator without having to re-impute the missing values
per bootstrap step.

The weighted bootstrap procedure has parallel steps as the DI procedure.
However, cautions should be taken when deriving the replicated DI
estimator. To preserve the imputation model in DI, we draw on the
idea of importance sampling \citep{geweke1989bayesian} to incorporate
the variability of the current replicated model parameter estimator
$\hat{\theta}^{(b)}$ and target parameter estimator $\hat{\tau}_{\DI}^{(b)}$
in each bootstrap iteration $b=1,\cdots,B$, where $B$ is the total
number of bootstrap replicates. A standard recommendation for the
total number of bootstrap replicates to get variance estimation is
$B=100$ \citep{boos2013essential}. In this way, one can approximate
the conditional expectation $\mathbb{E}\big\{\psi(Z_{i},\tau)|Z_{\text{obs},i},\hat{\theta}^{(b)}\big\}$
by a weighted sum as 
\begin{align*}
\mathbb{E}\big\{\psi(Z_{i},\tau)\mid Z_{\text{obs},i},\hat{\theta}^{(b)}\big\}\approx\sum_{m=1}^{M}w_{i}^{(m)}(\hat{\theta}^{(b)})\psi(Z_{i}^{*(m)},\tau),
\end{align*}
where the importance weights are computed as 
\begin{equation}
w_{i}^{(m)}(\hat{\theta}^{(b)})\propto\frac{f(Z_{i}^{*(m)}\mid Z_{\text{obs},i},\hat{\theta}^{(b)})}{f(Z_{i}^{*(m)}\mid Z_{\text{obs},i},\hat{\theta})},\label{importance_weight}
\end{equation}
with the constraint $\sum_{m=1}^{M}w_{i}^{(m)}(\hat{\theta}^{(b)})=1$
for all $i$.

To summarize, conduct the weighted bootstrap procedure in each iteration
as follows: 
\begin{description}
\item [{{Step 1}.}] Generate the i.i.d. bootstrap weights $u_{i}^{(b)}$
such that $\mathbb{E}(u_{i}^{(b)})=1,\mathbb{V}(u_{i}^{(b)})=1$ with
$u_{i}^{(b)}\geq0$ for each individual. Obtain the model parameter
estimator $\hat{\theta}^{(b)}$ by solving the estimating equations
\begin{equation}
\sum_{i=1}^{n}u_{i}^{(b)}\mathbb{E}\{s(Z_{i},\theta)\mid Z_{\text{obs},i}\}=0.\label{bootstrap_score}
\end{equation}
\item [{{Step 2}.}] Update the importance weights $w_{i}^{(m)}(\hat{\theta}^{(b)})$
such that it satisfies \eqref{importance_weight} with a constraint
$\allowbreak \sum_{m=1}^{M}w_{i}^{(m)}(\hat{\theta}^{(b)})=1$ for all $i$, under
the prespecified imputation settings such as Assumptions \ref{assump:j2r}--\ref{assump:washout}. 
\item [{{Step 3}.}] Obtain the DI estimator $\hat{\tau}_{\DI}^{(b)}$
by solving the estimating equations 
\begin{equation}
\sum_{i=1}^{n}\sum_{m=1}^{M}u_{i}^{(b)}w_{i}^{(m)}(\hat{\theta}^{(b)})\psi(Z_{i}^{*(m)},\tau)=0.\label{bootstrap_ee}
\end{equation}
\end{description}
Repeat Steps 1--3 $B$ times, and get the replication variance estimator
of the DI estimator as 
\[
\hat{\mathbb{V}}_{2}(\hat{\tau}_{\DI})=\frac{1}{B-1}\sum_{b=1}^{B}(\hat{\tau}_{\DI}^{(b)}-\hat{\tau}_{\DI})^{2}.
\]

\begin{remark}[Choice of the weight distribution] There are many
candidate distributions to generate the bootstrap weights $u_{i}^{(b)}$.
For example, one may try an exponential distribution with the rate
parameter 1 denoted as $\text{Exp}(1)$, or a discrete distribution
such as Poisson with parameter 1. The choice of the generated distribution
is not sensitive to the variance estimation. We adopt $\text{Exp}(1)$
in simulation studies. \end{remark}

Theorem \ref{theo:wb} shows the asymptotic validity of the above
weighted bootstrap method, with the proof in Section \ref{appen:thm3}
in the supplementary material.

\begin{theorem}\label{theo:wb} Under regularity conditions listed in
Sections \ref{appen:thm1} and \ref{appen:thm2} in the supplementary
material, with the bootstrap weights $u_{1}^{(b)},\cdots,u_{n}^{(b)}$
i.i.d. satisfying $\mathbb{E}(u_{i}^{(b)})=1,\mathbb{V}(u_{i}^{(b)})=1\text{ with }u_{i}^{(b)}\geq0$,
$\hat{\mathbb{V}}_{2}(\hat{\tau}_{\DI})$ is a consistent estimator
of $\mathbb{V}(\hat{\tau}_{\DI})$. \end{theorem}

\section{Simulation study \label{sec:simulation}}

We conduct simulation studies to assess the finite-sample validity
of our proposed framework using DI and weighted bootstrap for sensitivity
analyses in longitudinal clinical trials. Consider a clinical trail
with two groups and five visits. The baseline covariates $X$ are
generated from the standard normal distribution with $p=3$ dimensions.
For the longitudinal responses $Y_{i}$ of the $i$th individual,
we generate them independently from a multivariate normal distribution
as the full-data model \eqref{eq:general_model}, where for the control
and treatment group, the group-specific coefficients $\beta_{j1},\cdots,\beta_{j5}$
and covariance matrices $\Sigma^{(j)}$ for $j=1,2$ are presented
in Section \ref{appen:sim_result} in the supplementary material.

Consider the missing mechanism as MAR with a monotone missingness
pattern. More precisely, assume all the baseline responses are observed,
i.e. $R_{i1}=1$. For visit $k>1$, if $R_{ik-1}=0$, then $R_{il}=0$
for $l=k,\cdots,T$; otherwise let $R_{ik}\sim\text{Bernoulli}(\pi_{ik})$.
Model the observed probability $\pi_{ik}$ at visit $k>1$ as a function
of the observed information as $\text{logit}(\pi_{ik}|G_{i}=j)=\phi_{1j}+\phi_{2j}Y_{ik-1}$,
where $\phi_{1j},\phi_{2j}$ are tuning parameters for the observed
probabilities. We set $\phi_{11}=-3.2,\phi_{12}=-4.0,\phi_{21}=\phi_{22}=0.2$
to get the observed probabilities as 0.7865 and 0.7938 for control
and treatment group, respectively.

Different types of treatment effect estimands including the ATE, the
risk difference, and the QTE are assessed. When the primary interest
is the ATE, we use the ATE estimator motivated by ANCOVA since it
shows an increase of efficiency compared to the one obtained by directly
taking the sample average. When the risk difference is of interest,
we set a threshold $c=4.5$ and are interested in $\tau=P(Y_{iT}\geq4.5|G_{i}=2)-P(Y_{iT}\geq4.5|G_{i}=1)$.
When the QTE is of interest, we set $q=0.5$ to obtain the behavior
of median. We focus on the sensitivity analysis under J2R to describe
the deviation from MAR, which is consistent with our motivating example.
For illustration, we only present the result for the ATE under J2R.
Simulation results for sensitivity analyses under other sensitivity
models such as the RTB and washout imputation, along with other treatment
effect estimands under J2R are given in Section \ref{appen:sim_result}
in the supplementary material. We select the number of bootstrap replicates
$B=100$. Consider the sample size $N$ for each group to be the same
value ranging from $\{100,500,1000\}$ for each group, and the imputation
size $M$ ranging from $\{5,10,100\}$. Select $\text{Exp}(1)$ as
the generated distribution of the bootstrap weights.

We compare our proposed DI with MI in the simulation study. Rubin's
method and weighted bootstrap are applied to the MI and DI estimator
to get the variance estimation, respectively, under 1000 MC simulations.
The estimators are assessed using the point estimate (Point est),
MC variance denoted as true variance (True var), variance estimate
(Var est), relative bias of the variance estimate computed by $\Big[\mathbb{E}\big\{\mathbb{\hat{V}}(\hat{\tau})\big\}-\mathbb{V}(\hat{\tau})\Big]/\mathbb{V}(\hat{\tau})$,
coverage rate of $95\%$ confidence interval (CI) and mean CI length.
We choose the $95\%$ Wald CI estimated by $\big(\hat{\tau}-1.96\mathbb{\hat{V}}^{1/2}(\hat{\tau}),\hat{\tau}+1.96\mathbb{\hat{V}}^{1/2}(\hat{\tau})\big)$.

Table \ref{rbi_mean_table} shows the simulation result of the ATE
estimator. The point estimates from both DI and MI are closer to the
true value as the sample size increases, and their MC variances are
getting smaller. It indicates that the MI and DI estimators are consistent
and have comparable performances regarding the point estimation. The
efficiency of the estimator increases as the imputation size $M$
grows. For variance estimation, Rubin's method ends up overestimating
the true variance as expected, with a larger relative bias and a conservative
coverage rate. The variance estimate using weighted bootstrap in DI
is close to the true variance, with a well-controlled relative bias
and a coverage rate close to the empirical value. For other types
of estimands, the variance estimate of the QTE in MI and DI overestimates
the true variance when the sample size is small. But as the sample
size grows, the results from DI are much better, with the variance
estimate proximal to the true value. Therefore, a relatively large
sample size is suggested when estimating the QTE based on the simulation
results.

\begin{table}[!htbp]
\centering \caption{Simulation results under J2R of the ATE estimator. Here the true value
$\tau=1.5400$.}
%Point estimation (Point est), true variance (True var), variance estimates(Var est), relative bias of the variance estimator, coverage of interval estimates, mean $95\%$CI length using MI with Rubin's method and purposed FI with weighted bootstrap method.}
\scalebox{1}{ \resizebox{\textwidth}{!}{%
\begin{tabular}{ccccccccccccccccccc}
\hline 
 &  & \multicolumn{2}{c}{Point est} &  & \multicolumn{2}{c}{True var} &  & \multicolumn{2}{c}{Var est} &  & \multicolumn{2}{c}{Relative bias} &  & \multicolumn{2}{c}{Coverage rate} &  & \multicolumn{2}{c}{Mean CI length}\tabularnewline
 &  & \multicolumn{2}{c}{($\times10^{-2}$)} &  & \multicolumn{2}{c}{($\times10^{-2}$)} &  & \multicolumn{2}{c}{($\times10^{-2}$)} &  & \multicolumn{2}{c}{($\%$)} &  & \multicolumn{2}{c}{($\%$)} &  & \multicolumn{2}{c}{($\times10^{-2}$)}\tabularnewline
\cline{3-4} \cline{4-4} \cline{6-7} \cline{7-7} \cline{9-10} \cline{10-10} \cline{12-13} \cline{13-13} \cline{15-16} \cline{16-16} \cline{18-19} \cline{19-19} 
N & M & MI & DI &  & MI & DI &  & MI & DI &  & MI & DI &  & MI & DI &  & MI & DI\tabularnewline
\hline 
 & 5 & 150.84 & 150.36 &  & 14.86 & 14.60 &  & 20.92 & 14.70 &  & 40.85 & 0.74 &  & 97.90 & 94.90 &  & 178.76 & 149.61\tabularnewline
100 & 10 & 150.95 & 150.48 &  & 14.27 & 14.25 &  & 20.63 & 14.43 &  & 44.59 & 1.25 &  & 97.90 & 94.90 &  & 177.64 & 148.19\tabularnewline
 & 100 & 150.74 & 150.76 &  & 14.00 & 13.91 &  & 20.37 & 14.21 &  & 45.44 & 2.17 &  & 97.80 & 95.10 &  & 176.61 & 147.05\tabularnewline
\hline 
 & 5 & 153.52 & 153.13 &  & 3.06 & 3.07 &  & 4.08 & 3.03 &  & 33.44 & -1.25 &  & 98.00 & 94.50 &  & 79.09 & 67.99\tabularnewline
500 & 10 & 153.42 & 153.45 &  & 3.02 & 3.02 &  & 4.01 & 2.97 &  & 33.09 & -1.43 &  & 97.80 & 94.50 &  & 78.48 & 67.38\tabularnewline
 & 100 & 153.39 & 153.43 &  & 2.98 & 2.99 &  & 3.97 & 2.92 &  & 33.17 & -2.27 &  & 98.20 & 94.10 &  & 78.07 & 66.75\tabularnewline
\hline 
 & 5 & 154.27 & 154.06 &  & 1.46 & 1.47 &  & 2.03 & 1.52 &  & 39.23 & 3.48 &  & 97.60 & 94.90 &  & 55.84 & 48.14\tabularnewline
1000 & 10 & 154.09 & 154.06 &  & 1.45 & 1.44 &  & 2.00 & 1.49 &  & 38.63 & 3.54 &  & 97.70 & 94.80 &  & 55.45 & 47.70\tabularnewline
 & 100 & 154.12 & 154.11 &  & 1.43 & 1.42 &  & 1.98 & 1.46 &  & 38.55 & 2.99 &  & 97.90 & 94.20 &  & 55.18 & 47.26\tabularnewline
\hline 
\end{tabular}} } \label{rbi_mean_table}
\end{table}

Each type of estimands based on DI and MI has a comparable performance
of the point estimation under each prespecified sensitivity analysis
setting. The variance estimation using weighted bootstrap in DI outperforms
Rubin's MI combining method in all cases with much tolerable relative
biases for the variance estimation and better coverage probabilities.
The same interpretations of the results apply to other settings as
shown in Section \ref{appen:sim_result} in the supplementary material. 

\section{Return to the motivating example \label{sec:application}}

We apply our proposed DI framework to the motivating
example in Section \ref{sec:example} to uncover the treatment effect
in sensitivity analyses. Apart from comparing the performance between
MI with Rubin's rule and DI with the proposed weighted bootstrap for
the ATE and the risk difference, we also explore the QTE defined by
the quantile difference between the relative change of the HAMD-17
score in both the primary analysis under MAR and the sensitivity analysis
under J2R. \textcolor{black}{The three treatment effect estimands
are estimated through the estimating equations in Example \ref{example2}
after imputation. For the QTE, we do not limit it to one specific
quantile; instead, we present the estimated cumulative distribution
function (CDF) of the relative change from baseline for each group.
In the implementation of MI and DI, the imputation size $M=100$,
and the number of bootstrap replicates $B=100$. }

Tables \ref{tab:real_reg_nonpara} and \ref{tab:real_prop_nonpara}
present the analysis results with the use of MI and DI. The primary
analysis in the ``MAR'' rows in the tables indicates a parallel
performance of MI and DI, with similar point and variance estimates.
While MI with Rubin's variance estimator alters the study conclusion
under J2R in terms of the ATE and the risk difference, DI with the
weighted bootstrap procedure preserves a significant treatment effect
by producing smaller standard errors and narrower CIs. Applying DI
with the weighted bootstrap resolves the suspicion of the effectiveness
of the experimental drug, as it guarantees consistent variance estimators
of the treatment effect.

\begin{table}[!htbp]
\centering{}\centering \caption{Analysis of the HAMD-17 data of the ATE.\label{tab:real_reg_nonpara}}
\begin{tabular}{ccccccccc}
\hline 
 & \multicolumn{2}{c}{Point estimation} &  & \multicolumn{2}{c}{Standard error} &  & \multicolumn{2}{c}{P-value}\tabularnewline
\cline{2-3} \cline{3-3} \cline{5-6} \cline{6-6} \cline{8-9} \cline{9-9} 
Setting & \multicolumn{1}{c}{MI} & DI &  & \multicolumn{1}{c}{MI} & DI &  & \multicolumn{1}{c}{MI} & DI\tabularnewline
 & ($95\%$ CI) & ($95\%$ CI) &  &  &  &  &  & \tabularnewline
\hline 
MAR & -2.42 & -2.30 &  & 1.06 & 1.11 &  & 0.022 & 0.038\tabularnewline
 & (-4.49, -0.35) & (-4.47, -0.13) &  &  &  &  &  & \tabularnewline
J2R & -1.81 & -1.68 &  & 1.07 & 0.82 &  & 0.091 & 0.039\tabularnewline
 & (-3.91, 0.29) & (-3.28, -0.08) &  &  &  &  &  & \tabularnewline
RTB & -1.23 & -1.25 &  & 1.10 & 0.96 &  & 0.266 & 0.192\tabularnewline
 & (-3.39, 0.93) & (-3.13, 0.63) &  &  &  &  &  & \tabularnewline
Washout & -0.71 & -0.75 &  & 1.08 & 1.04 &  & 0.511 & 0.475\tabularnewline
 & (-2.83, 1.41) & (-2.79, 1.39) &  &  &  &  &  & \tabularnewline
\hline 
\end{tabular}
\end{table}

\begin{table}[!htbp]
\centering{}\centering \caption{Analysis of the HAMD-17 data of the risk difference.\label{tab:real_prop_nonpara}}
\begin{tabular}{ccccccccc}
\hline 
 & \multicolumn{2}{c}{Point estimation ($\%$)} &  & \multicolumn{2}{c}{Standard error ($\%$)} &  & \multicolumn{2}{c}{P-value}\tabularnewline
\cline{2-3} \cline{3-3} \cline{5-6} \cline{6-6} \cline{8-9} \cline{9-9} 
Setting & \multicolumn{1}{c}{MI} & DI &  & \multicolumn{1}{c}{MI} & DI &  & \multicolumn{1}{c}{MI} & DI\tabularnewline
 & ($95\%$ CI) & ($95\%$ CI) &  &  &  &  &  & \tabularnewline
\hline 
MAR & 15.64 & 15.53 &  & 7.48 & 6.89 &  & 0.037 & 0.024\tabularnewline
 & (0.98, 30.30) & (2.02, 29.03) &  &  &  &  &  & \tabularnewline
J2R & 12.81 & 12.78 &  & 7.44 & 5.95 &  & 0.085 & 0.032\tabularnewline
 & (-1.78, 27.40) & (1.11, 24.45) &  &  &  &  &  & \tabularnewline
RTB & 12.87 & 13.05 &  & 6.94 & 6.54 &  & 0.064 & 0.046\tabularnewline
 & (-0.73, 26.48) & (0.24, 25.86) &  &  &  &  &  & \tabularnewline
Washout & 8.40 & 8.42 &  & 7.17 & 6.74 &  & 0.241 & 0.211\tabularnewline
 & (-5.65, 22.45) & (-4.79, 21.63) &  &  &  &  &  & \tabularnewline
\hline 
\end{tabular}
\end{table}

\begin{figure}[!htbp]
\caption{The estimated CDF and QTE of relative change from baseline at last
time point via MI and DI in the primary analysis, accompanied by the
point-wise $95\%$ CI in dashed lines.\label{fig:primary}}

\includegraphics[width=0.49\textwidth]{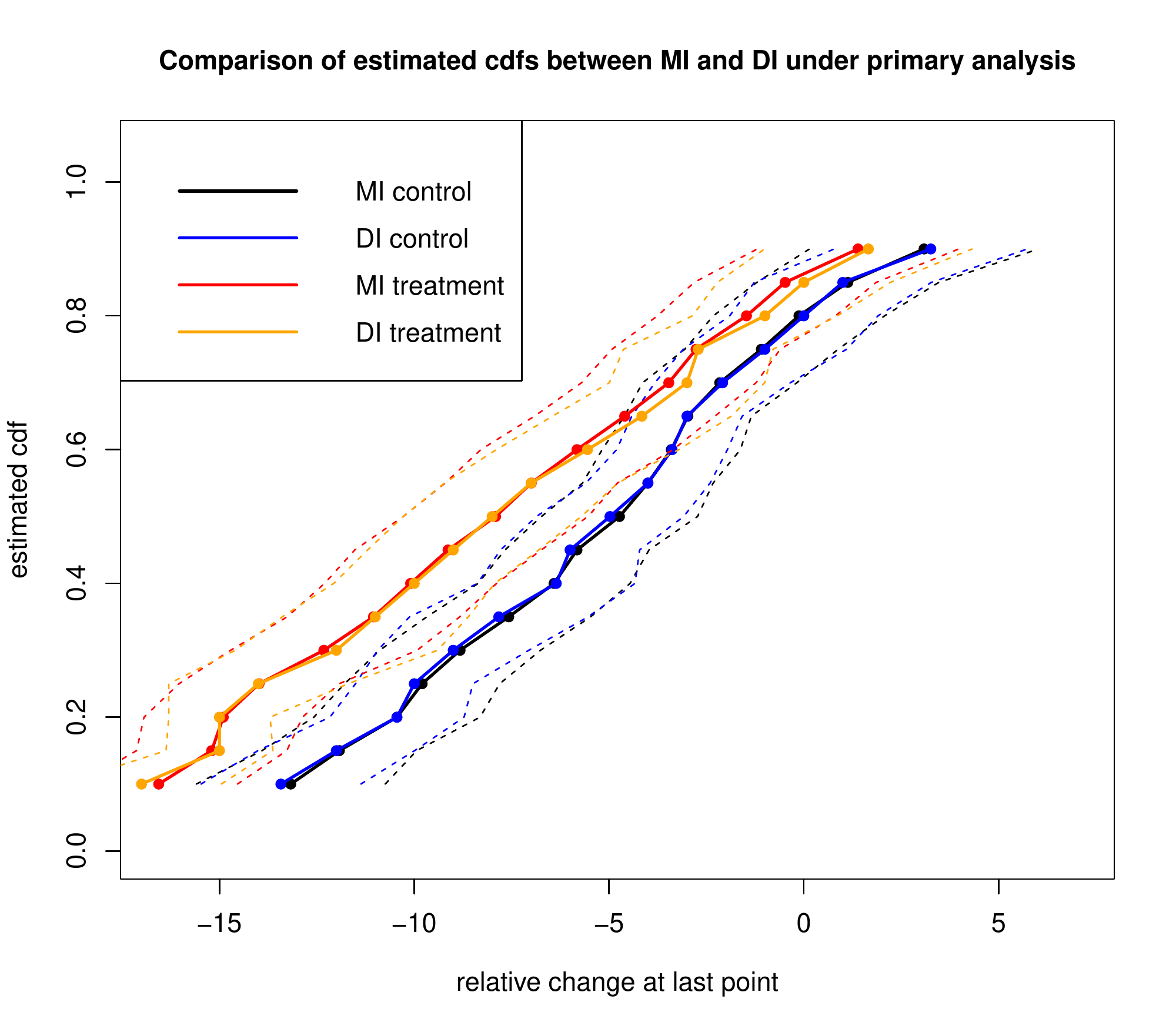}
\includegraphics[width=0.49\textwidth]{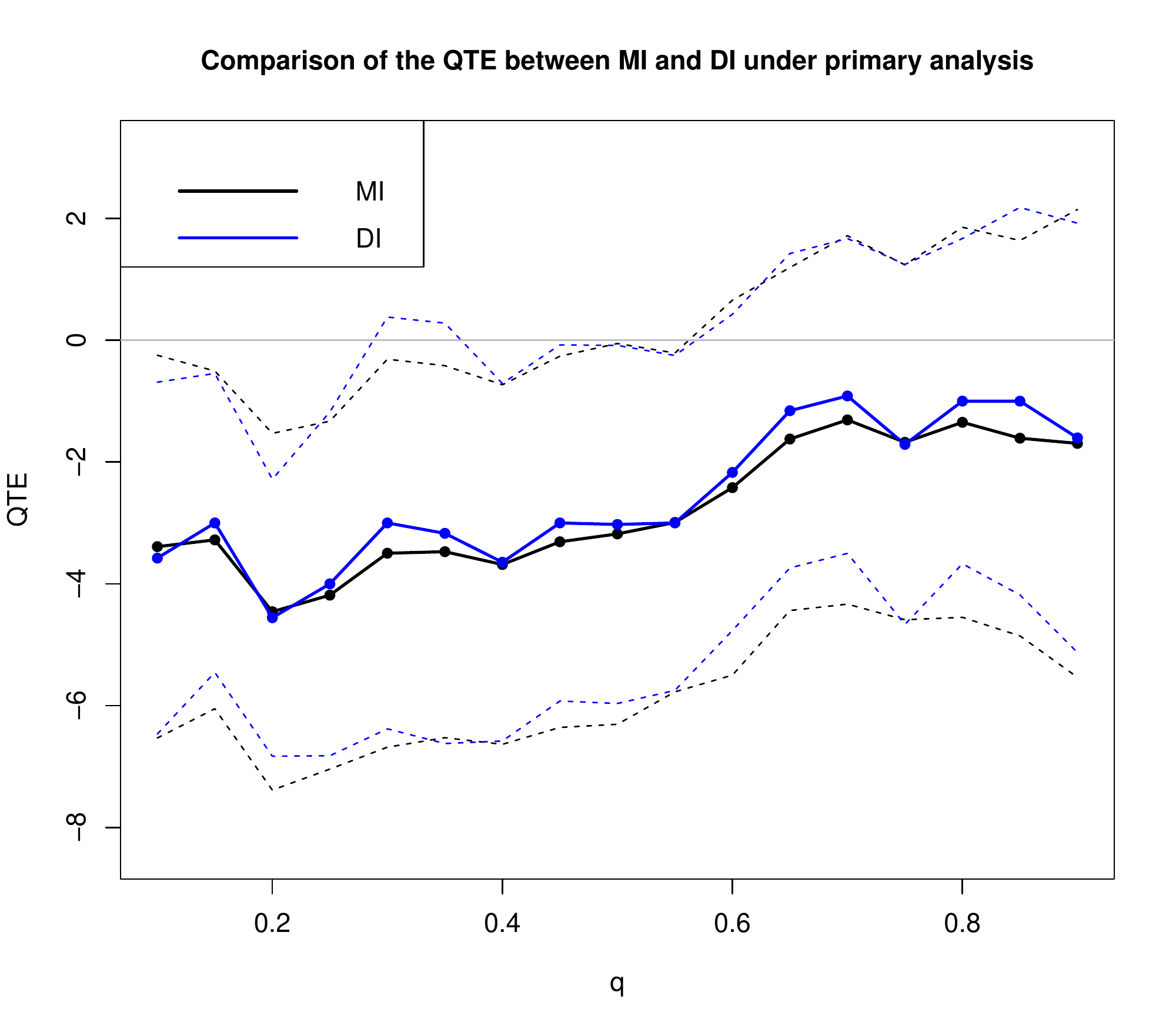} 
\end{figure}

\begin{figure}[!htbp]
\caption{The estimated CDF and QTE of relative change from baseline at last
time point via MI and DI in the sensitivity analysis under J2R, accompanied
by the point-wise $95\%$ CI in dashed lines.\label{fig:j2r}}

\includegraphics[width=0.49\textwidth]{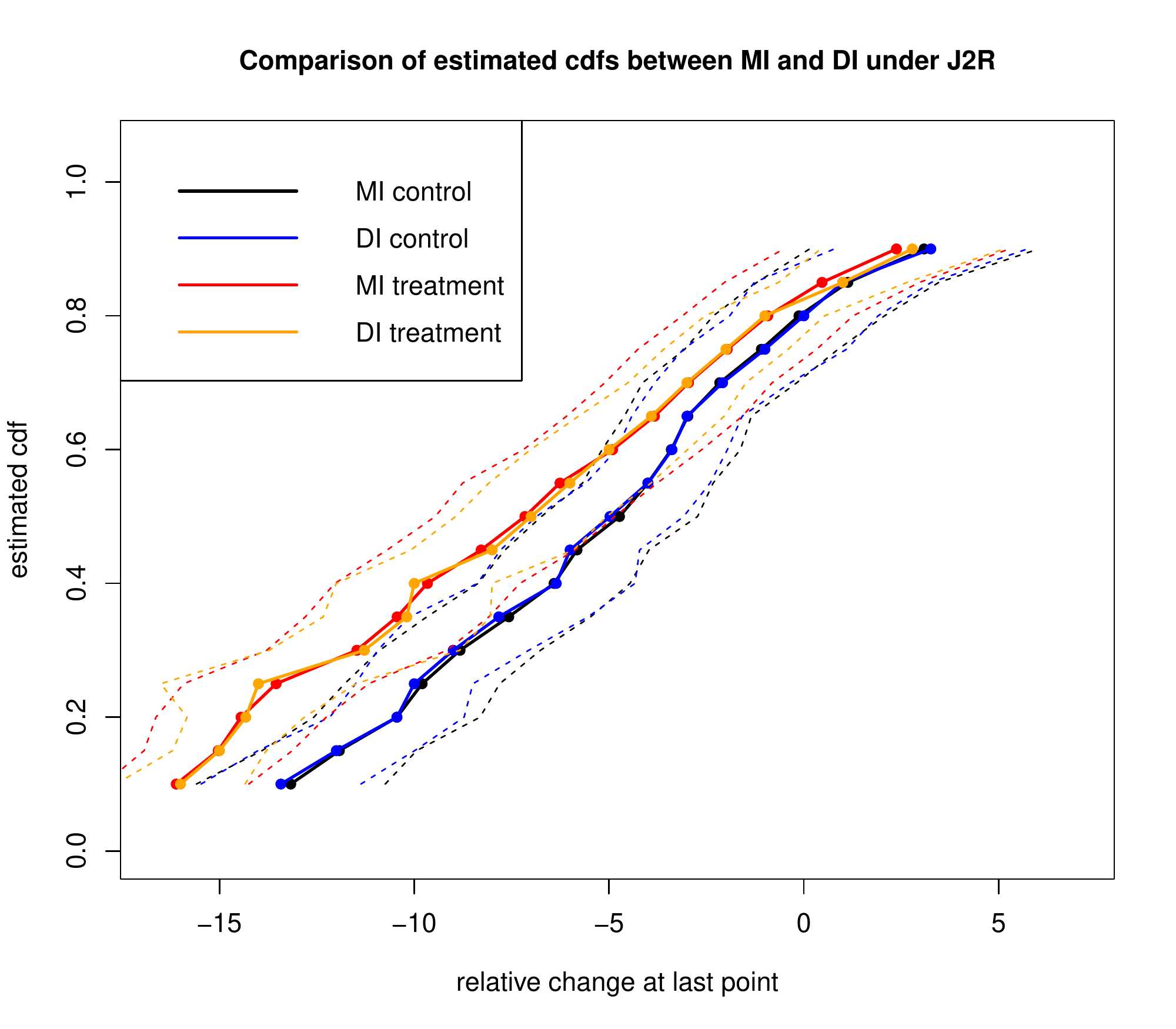}
\includegraphics[width=0.49\textwidth]{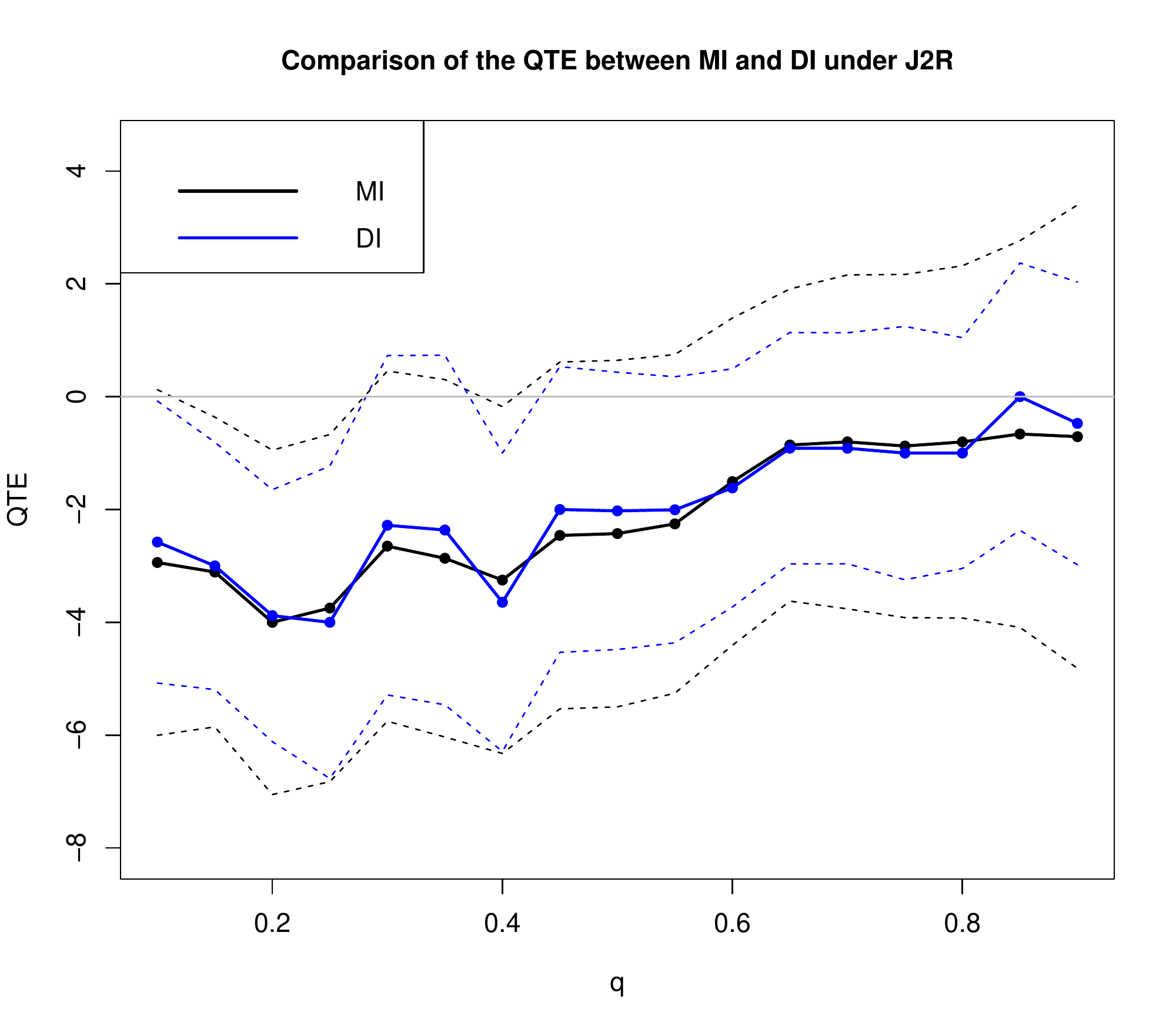} 
\end{figure}

To evaluate the distributional behavior of the data, we estimate the
CDF of the relative change of the HAMD-17 score at the last time point
for each group and the QTE as a function of $q$ denoted as the quantile
percentage under both MAR and J2R in Figures \ref{fig:primary} and
\ref{fig:j2r}. Similar to the results from the ATE and the risk difference,
the estimated CDF obtained from DI has a comparable shape as the one
obtained from MI, while the curve from DI has a narrower $95\%$ confidence
region in the sensitivity analysis compared to MI. A statistically
significant treatment effect is detected for patients in the lower
quantiles of the HAMD-17 score in both primary and sensitivity analyses.

Our general framework captures a comprehensive evaluation of the treatment
effect. With the use of DI, the experimental drug
reveals its significant benefit for curing depression in both the
primary analysis and the sensitivity analysis under J2R. If other
sensitivity models such as the RTB and washout imputation are assumed
in the trial protocol, we also provide the corresponding results in
Tables \ref{tab:real_reg_nonpara} and \ref{tab:real_prop_nonpara},
and in Section \ref{appen:real_result} in the supplementary material,
to illustrate a wide application of the proposed framework. Under
each sensitivity model, DI produces smaller standard errors and CIs
compared to MI. However, one should notice that the study conclusion
regarding the treatment effect changes with the prespecified sensitivity
models. Among those sensitivity analyses, only the result under J2R
using DI captures the same statistical significance of the ATE; the
result under washout imputation fails to show an improvement of the
treatment with respect to the risk difference. It suggests the potential impact of different missingness assumptions on the treatment effect of an experimental drug.  
The study statistician should 
carefully interpret the 
primary and sensitivity analyses results with 
investigators based on the missingness assumptions and additional clinical  knowledge of the drug.

\section{Conclusion \label{sec:conclude}}

In the paper, we propose DI using the idea of MC integration and establish
a unified framework for the sensitivity analysis in longitudinal clinical
trials to assess the impact of the MAR assumption in the primary analysis.
Our framework is flexible to accommodate various sensitivity models
and treatment effect estimands. We apply the proposed
DI with weighted bootstrap to an antidepressant longitudinal clinical
study and detect a statistically significant treatment effect in both
the primary and sensitivity analyses for the ATE, the risk difference,
and the QTE, overcoming the inefficiency and overestimation issue
from MI with Rubin's rule. The study result of the experimental drug
uncovers the significant improvement for curing depression. The DI
framework has a solid theoretical guarantee, with the avoidance of
re-imputation of the missing data in the variance estimation via weighted
bootstrap.

While we present DI under monotone missingness, DI is applicable to
handle intermittent missing values as long as the imputation model
of the missing values given the observed values is tractable. If direct
sampling from the target imputation model is difficult, one can resort
to alternative sampling strategies such as importance sampling, Metropolis--Hastings,
etc. We leave this topic as a future research direction.

Although we present the framework for sensitivity analyses using continuous
longitudinal outcomes, it is possible to extend the framework to the
cases of categorical or survival outcomes under some prespecified
imputation mechanisms. For example, \citet{tang2018controlled} modifies
the control-based imputation model for binary and ordinal responses
based on the generalized linear mixed model; \citet{yang2020smim}
adopts the $\delta$-adjusted and control-based imputation models
for survival outcomes in sensitivity analysis. With motivated treatment
effect estimands and suitable prespecified sensitivity assumptions,
our framework can handle sensitivity analyses for different types
of responses in clinical trials. \citet{guan2019unified} establish
a unified framework of MI via wild bootstrap for causal inference
in observational studies; extending the proposed DI inference to this
context is straightforward.

The proposed general framework for sensitivity analyses in longitudinal
clinical trials relies on parametric modeling assumptions in both
the imputation and analysis stages. This parametric setup is originated
from MI. In the future, we will develop DI under semiparametric and
nonparametric models as more flexible settings in sensitivity analyses.

\section*{Acknowledgements}
Yang is partially supported by the NSF grant DMS 1811245, NIA grant 1R01AG066883, and NIEHS grant 1R01ES031651.

\section*{Supplementary material}

The supplementary materials include the setup and proof for the theorems, additional simulation, and real data application results.

\bibliographystyle{Chicago}
\bibliography{DR_PSACE_ref}

\newpage{} 
\begin{center}
\textbf{\Large{}Supplementary material for "Sensitivity analysis in longitudinal clinical trials via distributional imputation" by Liu et al.}{\Large{} }{\Large\par}
\par\end{center}

\pagenumbering{arabic} %reset page counter to 1
\renewcommand*{\thepage}{S\arabic{page}}

\setcounter{lemma}{0} 
\global\long\def\thelemma{\textup{S}\arabic{lemma}}%
\setcounter{equation}{0} 
\global\long\def\theequation{S\arabic{equation}}%
\setcounter{section}{0} 
\global\long\def\thesection{S\arabic{section}}%
\global\long\def\thesubsection{S\arabic{section}.\arabic{subsection}}%
\setcounter{table}{0} 
\global\long\def\thetable{\textup{S}\arabic{table}}%
\setcounter{figure}{0} 
\global\long\def\thefigure{\textup{S}\arabic{figure}}%
\setcounter{thm}{0}
\global\long\def\thethm{\textup{S}\arabic{thm}}%
\setcounter{corollary}{0}
\global\long\def\thecorollary{\textup{S}\arabic{corollary}}%

This supplementary material contains technical details, additional simulation studies, and results for the real-data application. Section \ref{appen:thm} gives the regularity conditions and proof of the theorems. Section \ref{appen:sim_result} presents additional simulation results under RTB and washout imputation mechanisms. Section \ref{appen:real_data} contains additional analysis results regarding the QTE under RTB and washout imputation and the model diagnosis of the underlying modeling assumptions in the real-data application. 

\section{Setup and proof of the theorems}\label{appen:thm}
To emphasize that the DI estimator depends on sample size, we put subscript $n$ in $\hat\tau_{DI,n}$ for illustration in the following theorems.

\subsection{Setup and proof of Theorem \ref{consistency_thm}}\label{appen:thm1}

Let $s(Z, \theta), \psi(Z, \tau)$ satisfy the conditions listed in Section \ref{sec:notation}. Denote $$\bar s(\theta) = n^{-1} \sum_{i=1}^n \mathbb{E}\{ s(Z_i, \theta)|Z_{\text{obs},i} \},$$and $$\bar \psi_n(\tau, \theta):= n^{-1}\sum_{i=1}^n \mathbb{E}\big\{\psi(Z_i, \tau)|Z_{\text{obs},i}, \theta \big\}.$$ Assume the following regularity conditions:
 
\begin{enumerate}
\item[C1.] $\bar{s}(\hat{\theta})$ solves for equation \eqref{theta_score_eq},
and there exists a unique $\theta_{0}$ such that $\mathbb{E}\big\{ s(Z_{i},\theta_{0})\big\}=0$. 
\item[C2.] $\bar{s}(\theta)$ is dominated by an integrable function $g(Z_{\text{obs}})$
for all $Z_{\text{obs}}\subset\mathbb{R}^{d}$ and $\tau$ with respect
to the conditional distribution function $f(Z_{\text{obs}}|Z_{\text{mis}},\theta)$. 
\item[C3.] There exists a unique $\tau_{0}$ such that $\mathbb{E}(\psi(Z_{i},\tau_{0}))=0$. 
\item[C4.] $\bar{\psi}_{n}(\tau,\theta)$ is dominated by an integrable function
$g_{1}(Z_{\text{obs}})$ for all $Z_{\text{obs}}\subset\mathbb{R}^{d}$
and $\tau$ with respect to the conditional distribution function
$f(Z_{\text{obs}}|Z_{\text{mis}},\theta)$. 
\item[C5.] $\mathbb{V}\big\{\bar{\psi}_{n}(\tau,\theta)\big\}<\infty$. 
\end{enumerate}
Under regularity conditions listed above, we prove that the DI estimator
$\hat{\tau}_{DI,n}\xrightarrow{\mathbb{P}}\tau_{0}$ as the imputation
size $M\rightarrow\infty$ and sample size $n\rightarrow\infty$,
where $\hat{\tau}_{DI,n}$ solves for equation \eqref{fi_ee}.

\begin{proof}
First show the consistency of $\hat \theta$. Since $s(Z, \theta)$ is a continuous function for $\theta$ and a measurable function for $Z$, then $\bar s(\theta)$ is continuous for $\theta$ and measurable for $Z$. Based on the above observations and regularity condition C2, it satisfies the conditions for Theorem 2 in \cite{jennrich1969asymptotic}. Thus
\begin{equation*}
    \bar s(\theta) \xrightarrow{a.s.} \mathbb{E}\big\{s(Z_i, \theta)\big\} \text{ uniformly for } \forall \theta \in \Theta.
\end{equation*}

Then prove the consistency by contradiction. Suppose $\hat \theta$ does not converge to $\theta_0$ in probability, i.e., there exists a subsequence $\{\hat \theta_{n_k}\} \xrightarrow{\mathbb{P}} \theta_1 \neq \theta_0$. Therefore, for $\forall \varepsilon > 0$, by triangle inequality we have
\begin{align*}
    P\Big\{\big|\bar s(\hat \theta_{n_k}) - \mathbb{E}\big\{s(Z_i, \theta_1)\big\} \big| \geq \varepsilon\Big\} & \leq P\Big\{\big|\bar s(\hat \theta_{n_k}) - \mathbb{E}\big\{s(Z_i, \hat \theta_{n_k})\big\} \big| \geq \frac{\varepsilon}{2}\Big\}\\ &\quad +P\Big\{\big|\mathbb{E}\big\{s(Z_i, \hat \theta_{n_k})\big\} - \mathbb{E}\big\{s(z_i, \hat \theta_{1})\big\}\big| \geq \frac{\varepsilon}{2}\Big\}.
\end{align*}

By uniform convergence of $\bar s(\theta)$ for $\forall \theta$, the first term of the right hand-side converges to 0. By the continuity of $\mathbb{E}\big\{s(Z_i, \hat \theta)\big\}$ for $\theta$ and $\{\hat \theta_{k}\} \xrightarrow{\mathbb{P}} \theta_1 \neq \theta_0$, the second term of the right hand-side also converges to 0. Thus we have $
    \text{lim}_{n\rightarrow \infty} P\Big\{\big|\bar s(\hat \theta_{k}) - \mathbb{E}\big\{s(Z_i, \theta_1)\big\} \big| \geq \varepsilon\Big\} = 0
$, i.e., $\bar s(\hat \theta_{k}) \xrightarrow{\mathbb{P}} \mathbb{E}\big\{s(Z_i, \theta_1)\big\} \neq 0$. Since C1 demands the uniqueness of $\theta_0$ such that $\mathbb{E}\big\{s(Z_i, \theta_0)\big\} = 0$ and $\hat \theta$ as the solution for equation \eqref{theta_score_eq} which means $\bar s(\hat \theta_{k}) \xrightarrow{\mathbb{P}} 0$, contradicting to what is proved above. Therefore $\hat \theta \xrightarrow{\mathbb{P}} \theta_0$. 

Denote $\bar \psi(\tau, \theta):= \mathbb{E}\big\{\psi(Z_i, \tau) | \theta \big\}$. Note that $\psi(Z,\tau)$ is continuous with respect to $\tau$, and a measurable function of $z$ for each $\tau$, then $\bar \psi_n(\tau, \theta)$ and $\bar \psi(\tau, \theta)$ are continuous functions to $\tau$. Also, $\bar \psi_n(\tau, \theta)$ is a measurable function of $z_{\text{obs}}$ for each $\tau$. By continuous mapping theorem, for $\forall \tau$, $\bar \psi_n(\tau, \hat \theta) - \bar \psi_n(\tau, \theta_0) \xrightarrow{\mathbb{P}} 0$ and $\bar \psi(\tau, \hat \theta) \xrightarrow{\mathbb{P}} \bar \psi(\tau, \theta_0)$. 

Note that $Z^{*(m)}_{i}$ are generated via Monte Carlo integration. By C5, $\mathbb{V}\Big[ \mathbb{E}\big\{\psi(Z_i,\tau)|Z_{\text{obs},i}, \theta \big\}\Big]$ is finite. From the asymptotic theory of Monte Carlo approximation in \cite{lepage1978new}, we have
\begin{align}\label{eq:mc_integration}
    \frac{1}{n}\frac{1}{M}\sum_{i=1}^n \sum_{m = 1}^M \psi(Z^{*(m)}_{i},\tau) &= \frac{1}{n}\sum_{i=1}^n \Big[\mathbb{E}\big\{\psi(Z_i, \tau)|Z_{\text{obs},i}, \theta \big\} + O_p(M^{-1/2}) \Big] \nonumber \\
    & = \bar \psi_n(\tau, \theta) + O_p(M^{-1/2}).
\end{align}

Since $\hat \tau_{DI,n}$ solves for equation \eqref{fi_ee}, as $M \rightarrow \infty$, we have
\begin{equation*}
    \bar \psi_n(\hat \tau_{DI,n}, \hat \theta) \xrightarrow{\mathbb{P}} 0.
\end{equation*}

Regularity condition C4, the assumptions for $\psi(Z, \tau)$ and the consistency of $\hat \theta$ satisfy the conditions for Theorem 2 in \cite{jennrich1969asymptotic}, then as $n \rightarrow \infty$
\begin{equation*}
    \bar \psi_n(\tau, \hat \theta) \xrightarrow{a.s.} \bar \psi(\tau, \theta_0) \text{ uniformly for } \forall \tau \in \Omega.
\end{equation*}

Then again prove the consistency by contradiction. Suppose $\hat \tau_{DI,n}$ does not converge to $\tau_0$ in probability, i.e., there exists a subsequence $\{\hat \tau_{n_k}\} \xrightarrow{\mathbb{P}} \tau_1 \neq \tau_0$. Therefore, for $\forall \varepsilon > 0$, by triangle inequality we have
\begin{align*}
    P\Big\{\big|\bar \psi_n(\hat \tau_{n_k}, \hat \theta) - \bar \psi(\tau_1, \theta_0)\big| \geq \varepsilon\Big\} &\leq P\Big\{\big|\bar \psi_n(\hat \tau_{n_k}, \hat \theta) - \bar \psi(\hat \tau_{n_k}, \theta_0)\big| \geq \frac{\varepsilon}{2}\Big\}  \\& \quad+ P\Big\{\big|\bar \psi(\hat \tau_{n_k}, \theta_0) - \bar \psi(\tau_1, \theta_0)\big| \geq \frac{\varepsilon}{2}\Big\}.
\end{align*}

By uniform convergence of $\bar \psi_n(\tau, \hat \theta)$ for $\forall \tau$ and consistency of $\hat \theta$, the first term of the right hand-side converges to 0. By the continuity of $\bar \psi(\tau, \theta_0)$ and $\{\hat \tau_{n_k}\} \xrightarrow{\mathbb{P}} \tau_1 \neq \tau_0$, the second term of the right hand-side also converges to 0. Thus we have $
    \text{lim}_{n\rightarrow \infty} P\Big\{\big|\bar \psi_n(\hat \tau_{n_k}, \hat \theta) - \bar \psi(\tau_1, \theta_0)\big| \geq \varepsilon\Big\} = 0
$, i.e., $\bar \psi_n(\hat \tau_{n_k}, \hat \theta) \xrightarrow{\mathbb{P}} \bar \psi(\tau_1, \theta_0) \neq 0$ since regularity condition C4 demands the uniqueness of $\tau_0$ such that $\bar \psi(\tau, \theta_0) = 0$, which contradicts the fact that $\bar \psi_n(\hat \tau_{n_k}, \hat \theta) \xrightarrow{\mathbb{P}} 0$ as proved above. Therefore $\hat \tau_{DI,n} \xrightarrow{\mathbb{P}} \tau_0$. 

\end{proof}

\subsection{Setup and proof of Theorem \ref{theo:asym_normal}}\label{appen:thm2}

Let $s(Z, \theta), \psi(Z, \tau)$ satisfy the conditions listed in Section \ref{sec:notation}. Use the same notations in Theorem \ref{consistency_thm}, and assume the regularity conditions C1 - C5 hold. Add the additional regularity conditions:
\begin{enumerate}
  \item[C6.] The solution $\hat \theta$ in \eqref{theta_score_eq} satisfies $\hat \theta - \theta_0 = O_p(n^{-1/2})$.
  \item[C7.] The partial derivatives of $\bar s(\theta), \bar \psi(\tau, \theta)$ with respect to $\theta$ exist and are continuous around $\theta_0$ almost everywhere. The second derivatives of $\bar s(\theta), \bar \psi(\tau, \theta)$ with respect to $\theta$ are continuous and dominated by some integrable functions.
  \item[C8.] The partial derivative of $\bar s(\theta)$ satisfies
  \begin{equation*}
      \lVert \frac{\partial \bar s(\theta)}{\partial \theta} - \mathbb{E}\big\{\frac{\partial \bar s(\theta)}{\partial \theta}\big\} \rVert \xrightarrow{\mathbb{P}} 0 \text{ uniformly in } \theta,
  \end{equation*}
  and $\mathbb{E}\big\{\partial \bar s(\theta)/\partial \theta \big\}$ is continuous and nonsingular at $\theta_0$. The partial derivative of $\bar \psi(\tau, \theta)$ with respect to $\theta$ satisfies
  \begin{equation*}
      \lVert \frac{\partial \bar \psi(\tau, \theta)}{\partial \theta} - \mathbb{E}\big\{\frac{\partial \bar \psi(\tau, \theta)}{\partial \theta}\big\} \rVert \xrightarrow{\mathbb{P}} 0 \text{ uniformly in } \theta,
  \end{equation*}
  and $\mathbb{E}\big\{\partial \bar \psi(\tau, \theta)/\partial \theta \big\}$ is continuous with respect to $\theta$ at $\theta_0$.
  \item[C9.] There exists $a > 0$ such that $\mathbb{E}\big\{\psi(Z_i, \tau)^{2+a} \big\} < \infty$ and $\mathbb{E}\big\{s_{j}(Z_i, \theta)^{2+a} \big\} < \infty$ where $s_{j}(Z_i, \theta) = \partial f(Z_i,\theta)/\partial \theta_j$ for $j = 1, \cdots, q$ and $\theta_j$ is the $j$th element of $\theta$.
  \item[C10.] $\psi(Z, \tau)$ and its first two derivatives with respect to $\tau$ exist for all $y \in \mathbb{R}$ and $\tau$ in a neighborhood of $\tau_0$, where $\mathbb{E} \big\{\psi(Z_i, \tau_0)\big\} = 0$. 
  \item[C11.] For each $\tau$ in a neighborhood of $\tau_0$, there exists an integrable function $g_2(Z)$ such that for all $j,k \in \{1, \cdots, d\}$ and all $Z$, 
  \begin{equation*}
      \big|\frac{\partial^2}{\partial \tau_j \tau_k}\psi_l(Z, \tau) \big| \leq g_2(Z).
  \end{equation*}
  \item[C12.] $A(\tau_0, \theta_0) = \mathbb{E}\Big[\partial \psi(Z_i, \tau_0)/\partial \tau + \big\{\partial \Gamma(\tau_0, \theta_0)/\partial \tau \big\}\bar s_i(\theta_0)  \Big]$ exists and is nonsingular.
  \item[C13.] $B(\tau_0, \theta_0) = \mathbb{V}\big\{\psi_i^{*}(\tau_0, \theta_0) + \Gamma(\tau_0, \theta_0)^{\intercal}\bar s_i(\theta_0)\big\}$ exists and is finite, where
  $\psi_i^{*}(\tau, \theta)= M^{-1}\sum_{m = 1}^M  \psi(Z^{*(m)}_{i},\tau)$, $\bar s_i(\theta) = \mathbb{E}\{s(Z_i, \theta) | Z_{\text{obs},i}\}$, $\Gamma(\tau, \theta_0) = \mathbf{I}_{\text{obs}}(\theta_0)^{-1} \mathbf{I}_{\psi, \text{mis}}(\tau, \theta_0)$. Here $\mathbf{I}_{\text{obs}}(\theta) = \mathbb{E}\{-\partial \bar s_i(\theta)/\partial \theta\}$, $\mathbf{I}_{\psi, \text{mis}}(\tau, \theta) = \mathbb{E}[\{s(Z_i, \theta) - \bar s_i(\theta)\}\psi(Z_{i},\tau)]$.
\end{enumerate}

Under regularity conditions, as the imputation size $M \rightarrow \infty$, we prove that
\begin{equation*}
    \sqrt{n}(\hat \tau_{DI,n} - \tau_0) \xrightarrow{d} \mathcal{N}(0, A(\tau_0, \theta_0)^{-1}B(\tau_0, \theta_0)\{A(\tau_0, \theta_0)^{-1}\}^{\intercal}).
\end{equation*}

\begin{proof}
In Theorem 1, we prove that $\hat \theta \xrightarrow{\mathbb{P}} \theta_0$. Consider a Taylor series expansion of $\bar s(\hat \theta)$ around $\theta_0$, there exists $\Tilde{\theta}$ that is between $\hat \theta$ and $\theta_0$, such that

\begin{equation*}
   \bar s(\hat \theta) = \bar s(\theta_0) + \frac{\partial \bar s_i(\theta_0)}{\partial \theta^{\intercal}}(\hat \theta - \theta_0) + \frac{1}{2} (\hat \theta - \theta_0)^{\intercal} \big\{\frac{\partial^2 \bar s(\Tilde{\theta})}{\partial \theta \partial \theta^{\intercal}}\big\} (\hat \theta - \theta_0).
\end{equation*}

Note that by C7, $\partial^2 \bar s(\Tilde{\theta})/(\partial \theta \partial \theta^{\intercal}) = O_p(1)$, thus by C6, the second term is $o_p(n^{-1/2})$. Since $\hat \theta$ solves for \eqref{theta_score_eq}, we have
\begin{align}\label{eq:theta_taylor}
    \hat \theta - \theta_0 & = \big\{-\frac{\partial \bar s( \theta_0)}{\partial \theta^{\intercal}}\big\}^{-1} \big\{\bar s(\theta_0) \big\} + o_p(n^{-1/2}) \nonumber \\
    & = \big\{-\frac{1}{n}\sum_{i=1}^n \frac{\partial \bar s_i( \theta_0)}{\partial \theta^{\intercal}}\big\}^{-1} \big\{\frac{1}{n}\sum_{i=1}^n \bar s_i(\theta_0) \big\} + o_p(n^{-1/2}).
\end{align}

Consider a Taylor series expansion of $\bar{\psi}_n(\tau, \hat \theta)$ with respect to $\theta$ around $\theta_0$, there exists $\Tilde{\theta}$ such that it is between $\theta_0$ and $\hat \theta$ and satisfies
\begin{equation*}
    \bar{\psi}_n(\tau, \hat \theta) - \bar{\psi}_n(\tau, \theta_0) = \frac{\partial \bar{\psi}_n(\tau, \theta_0)}{\partial \theta^{\intercal}} (\hat \theta - \theta_0) + \frac{1}{2}(\hat \theta - \theta_0)^{\intercal}\big\{\frac{\partial^2 \bar{\psi}_n(\tau; \Tilde{\theta})}{\partial \theta\partial \theta^{\intercal}} \big\}(\hat \theta - \theta_0) + o_p(n^{-1/2}).
\end{equation*}

By C7, $\partial^2 \bar{\psi}_n(\tau; \Tilde{\theta})/(\partial \theta \partial \theta^{\intercal}) = O_p(1)$. Therefore by C6, the second term of the right-hand side is $o_p(n^{-1/2})$. We proceed to compute the partial derivative $\partial \bar{\psi}_n(\tau, \theta_0)/\partial \theta^{\intercal}$ as follows.
\begin{align*}
    \frac{\partial \bar{\psi}_n(\tau, \theta_0)}{\partial \theta} & = \frac{1}{n}\sum_{i=1}^n \int \frac{\partial f(Z_i|Z_{\text{obs},i}, \theta_0)}{\partial \theta} \psi(Z_i, \tau) dZ_i \\
    & = \frac{1}{n}\sum_{i=1}^n \int \frac{\partial \log \big\{f(Z_i|Z_{\text{obs},i}, \theta_0)\big\}}{\partial \theta}f(Z_i|z_{\text{obs},i}, \theta_0) \psi(Z_i, \tau) dZ_i \\
    & = \frac{1}{n}\sum_{i=1}^n \int \Big[\frac{\partial \log \big\{f(Z_i| \theta_0)\big\}}{\partial \theta} - \frac{\partial \log \big\{f(Z_{\text{obs}, i}| \theta_0)\big\}}{\partial \theta} \Big]f(Z_i|Z_{\text{obs},i}, \theta_0) \psi(Z_i, \tau) dZ_i \\
    & = \frac{1}{n}\sum_{i=1}^n \int \Big[s(Z_i, \theta_0) - \frac{\partial \log \big\{f(Z_{\text{obs}, i}| \theta_0)\big\}}{\partial \theta}\Big]f(Z_i|Z_{\text{obs},i}, \theta_0) \psi(Z_i, \tau) dZ_i \\
    & = \frac{1}{n}\sum_{i=1}^n \mathbb{E}\Big[ \big\{s(Z_i, \theta_0) - \bar s_i(\theta_0)\big\}\psi(Z_i, \tau)\big| Z_{\text{obs},i}, \theta_0 \Big].
\end{align*}
Note that in the first line of the equations above, we interchange the integral and derivative since the support of $f(Z, \theta_{0})$ is free of $\theta_{0}$. The subsequent lines are derived from basic analysis.

Therefore, we can rewrite the Taylor expansion of $\bar \psi_n(\tau, \hat \theta)$ around $\theta_0$ based on the last equation and \eqref{eq:theta_taylor} as
\begin{align*}
    \bar{\psi}_n(\tau, \hat \theta) - \bar{\psi}_n(\tau, \theta_0) & = \Bigg(\frac{1}{n}\sum_{i=1}^n \mathbb{E}\Big[\big\{s(Z_i, \theta_0) - \bar s_i(\theta_0)\big\}\psi(Z_i, \tau) \big| z_{\text{obs},i}, \Big]\Bigg) \\
    &\quad \times \big\{-\frac{1}{n}\sum_{i=1}^n \frac{\partial \bar s_i( \theta_0)}{\partial \theta^{\intercal}}\big\}^{-1} \big\{\frac{1}{n}\sum_{i=1}^n \bar s_i(\theta_0) \big\}    + o_p(n^{-1/2}).
\end{align*}

By weak law of large number and C8, $n^{-1}\sum_{i=1}^n \mathbb{E}\Big[\big\{s(Z_i, \theta_0) - \bar s_i(\theta_0)\big\}\psi(Z_i, \tau) \big| Z_{\text{obs},i}, \theta_0 \Big] = \mathbb{E}\Big[\big\{s(Z_i, \theta_0) - \bar s_i(\theta_0)\big\}\psi(Z_i, \tau) \Big] + o_p(1)$, and $$\big\{n^{-1}\sum_{i=1}^n \partial \bar s_i(\theta_0)/\partial \theta^{\intercal}\big\}^{-1} = \Big[\mathbb{E}\big\{\partial \bar s_i(\theta_0)/\partial \theta^{\intercal}\big\}\Big]^{-1} + o_p(1).$$ Denote $$\Gamma(\tau, \theta_0) := I_{\text{obs}}(\theta_0)^{-1}I_{\psi, \text{mis}}(\tau, \theta_0) = \Big[\mathbb{E}\big\{-\frac{\partial \bar s_i(\theta_0)}{\partial \theta^{\intercal}}\big\}\Big]^{-1}\mathbb{E}\Big[\big\{s(Z_i, \theta_0) - \bar s_i(\theta_0)\big\}\psi(Z_i, \tau) \Big].$$
Then we have
\begin{align*}
    \bar{\psi}_n(\tau, \hat \theta) - \mathbb{E}\big\{\psi(Z_i, \tau) \big\} & = \frac{1}{n}\sum_{i=1}^n\Big[\Gamma(\tau, \theta_0)^{\intercal}\bar s_i(\theta_0) + \mathbb{E}\big\{\psi(Z_i, \tau)|Z_{\text{obs,i}}, \theta_0\big\} \Big] + o_p(n^{-1/2}) \\
    & = \frac{1}{n}\sum_{i=1}^n\Big[\Gamma(\tau, \theta_0)^{\intercal}\bar s_i(\theta_0) + \psi^*(\tau, \theta_0) \Big] + o_p(n^{-1/2}) + O(M^{-1/2}).
\end{align*}

Thus for any $\tau$, as the imputation size $M \rightarrow \infty$ and sample size $n \rightarrow \infty$,
\begin{equation}\label{asym_normal}
    \sqrt{n}\Big[\bar{\psi}_n(\tau; \hat \theta) - \mathbb{E}\big\{\psi(Z_i, \tau) |\theta_0 \big\}\Big] \xrightarrow{d} \mathcal{N}(0, B(\tau, \theta_0)).
\end{equation}

Denote
\begin{align*}
    \Tilde{\psi}_n(\tau, \theta) = \frac{1}{nM}\sum_{i=1}^n \sum_{m = 1}^M \psi(Z^{*(m)}_{i},\tau) + \frac{1}{n}\sum_{i=1}^n \Gamma(\tau, \theta)^{\intercal}\bar s_i(\theta).
\end{align*}

Consider a Taylor series expansion of $\Tilde{\psi}(\hat \tau_{DI,n}, \hat \theta)$ around $\tau_0$ in a component-wise way. For $l = 1, \cdots, q$ representing the $l$th component of a vector, there exists $\Tilde{\tau}_l^*$ between $\hat \tau_{DI,n_l}$ and $\tau_{0l}$, such that
\begin{equation*}
    \Tilde{\psi}_{n}(\hat \tau_{DI,n}, \hat \theta) = \Tilde{\psi}_{n_l}(\tau_{0}, \hat \theta) + \frac{\partial \Tilde{\psi}_{n_l}(\tau_{0}, \hat \theta)}{\partial \tau_{l}} (\hat \tau_{DI,n} - \tau_{0}) + \frac{1}{2}(\hat \tau_{DI,n} - \tau_{0})^{\intercal}\frac{\partial^2 \Tilde{\psi}_{n_j}(\Tilde{\tau}_l^*, \hat \theta)}{\partial \tau_l^2}(\hat \tau_{DI,n} - \tau_{0}).
\end{equation*}

Stack the above $q$ equations together, 
\begin{align*}
    \bar{\psi}_{n}(\hat \tau_{DI,n}, \hat \theta) = \bar{\psi}_{n}(\tau_0, \hat \theta) + \big\{\frac{\partial \bar \psi_{n}(\tau_0, \hat \theta)}{\partial \tau^{\intercal}} + \frac{1}{2}(\hat \tau_{DI,n} - \tau_0)^{\intercal} \Tilde{Q}^* \big\}(\hat \tau_{DI,n} - \tau_0),
\end{align*}
where $\Tilde{Q}^*$ is a matrix with the $j$th row vector equals to $\partial^2 \Tilde{\psi}_{n_l}(\Tilde{\tau}_l^*, \hat \theta)/\partial \tau_l^2$. From C9, each row vector is $O_p(1)$. Thus $(\hat \tau_{DI,n} - \tau_0)^{\intercal} \Tilde{Q}^* = o_p(1)$ by Theorem 1. 

By weak law of large number and C6, 
\begin{align*}
    -\frac{\partial \Tilde \psi_{n}(\tau_0, \hat \theta)}{\partial \tau} & \xrightarrow{\mathbb{P}} \mathbb{E}\Big[\frac{\partial \mathbb{E}\big\{\psi(Z_i, \tau_0)|Z_{\text{obs},i} \big\}}{\partial \tau} + \frac{\partial \Gamma(\tau_0, \theta_0)}{\partial \tau}\mathbb{E}\big\{ s(Z_i, \theta_0)|Z_{\text{obs},i}\big\}  \Big] \\
    & = \mathbb{E}\Big[\partial \psi(Z_i, \tau_0)/\partial \tau + \big\{\partial \Gamma(\tau_0, \theta_0)/\partial \tau \big\}\bar s_i(\theta_0)  \Big] =  A(\tau_0, \theta_0).
\end{align*}
By C10, $A(\tau_0, \theta_0)$ is nonsingular. By C8, $ \Tilde{\psi}_{n}(\hat \tau_{DI,n}, \hat \theta) = o_p(n^{-1/2})$. Thus we can re-express the stack form of the equations as
\begin{equation*}
    \hat \tau_{DI,n} - \tau_0 = A(\tau_0, \theta_0)^{-1}\Tilde{\psi}_{n}(\tau_0, \hat \theta) + o_p(n^{-1/2}).
\end{equation*}
Let $\tau = \tau_0$ in \eqref{asym_normal}, and by C10,$\sqrt{n}\Tilde{\psi}_{n}(\tau_0, \hat \theta) \xrightarrow{d} \mathcal{N}\big(0, B(\tau_0, \theta_0)\big)$.
Then by Slutsky's theorem, we have
\begin{equation*}
    \sqrt{n}(\hat \tau_{DI,n} - \tau_0) \xrightarrow{d} \mathcal{N}\big(0, A(\tau_0, \theta_0)^{-1}B(\tau_0, \theta_0)\{A(\tau_0, \theta_0)^{-1}\}^{\intercal}\big).
\end{equation*}
\end{proof}

\subsection{Proof of Theorem \ref{theo:wb}}\label{appen:thm3}

\begin{proof}
Denote $w^{(m)}_i(\hat \theta) = M^{-1}$, and define
\begin{equation*}
    \bar \psi^{*(b)}_n (\tau, \theta) = \frac{1}{n}\sum_{i=1}^n u^{(b)}_i \sum_{m=1}^M w^{(m)}_i(\theta)\psi(Z_{i}^{*(m)}, \tau).
\end{equation*}

Given the complete data, $\hat \theta^{(b)} \xrightarrow{\mathbb{P}} \hat \theta$ by the same argument from Theorem 1. Consider a Taylor series expansion of $n^{-1}\sum_{i = 1}^n u_{i}^{(b)} \bar s_i(\hat \theta^{(b)}) = n^{-1}\sum_{i = 1}^n u_{i}^{(b)}\mathbb{E}\{s(Z_i, \hat \theta^{(b)})| Z_{\text{obs},i}\}$ around $\hat \theta$, there exists $\Tilde{\theta}^{(b)}$ that is between $\hat \theta^{(b)}$ and $\hat \theta$, such that
\begin{align*}
   \frac{1}{n}\sum_{i = 1}^n u_{i}^{(b)} \bar s_i(\hat \theta^{(b)}) &= \frac{1}{n}\sum_{i = 1}^n u_{i}^{(b)} \bar s_i(\hat \theta) + \frac{1}{n}\sum_{i=1}^n u_{i}^{(b)} \frac{\partial \bar s_i(\hat \theta)}{\partial \theta^{\intercal}}(\hat \theta^{(b)} - \hat \theta) \\& \quad + \frac{1}{2} (\hat \theta^{(b)} - \hat \theta)^{\intercal} \big\{\frac{1}{n}\sum_{i=1}^n u_{i}^{(b)} \frac{\partial^2 \bar s_i(\Tilde{\theta}^{(b)})}{\partial \theta \partial \theta^{\intercal}}\big\} (\hat \theta^{(b)} - \hat \theta).
\end{align*}

Note that by C5, $\sum_{i=1}^n u_{i}^{(b)} \big\{\partial^2 \bar s_i(\Tilde{\theta}^{(b)})/(\partial \theta \partial \theta^{\intercal})\big\} = O_p(1)$, thus $$(\hat \theta^{(b)} - \hat \theta)^{\intercal} \Big[n^{-1}\sum_{i=1}^n u_{i}^{(b)} \big\{\partial^2 \bar s_i(\Tilde{\theta}^{(b)})/(\partial \theta \partial \theta^{\intercal})\big\}\Big] (\hat \theta^{(b)} - \hat \theta) = o_p(n^{-1/2}).$$ Note that $\hat \theta^{(b)}$ solves for \eqref{bootstrap_score}, then

\begin{equation}\label{boot_thetab}
    \hat \theta^{(b)} - \hat \theta = \big\{\frac{1}{n}\sum_{i=1}^n u_{i}^{(b)} \frac{\partial \bar s_i(\hat \theta)}{\partial \theta^{\intercal}}\big\}^{-1} \big\{\frac{1}{n}\sum_{i = 1}^n u_{i}^{(b)} \bar s_i(\hat \theta) \big\}.
\end{equation}

Also consider a Taylor series expansion of $\bar \psi^{*(b)}_n (\tau, \hat \theta^{(b)})$ around $\hat \theta$, there exists $\Tilde{\theta}^{*(b)}$ that is between $\hat \theta^{(b)}$ and $\hat \theta$, such that
\begin{equation*}
    \bar \psi^{*(b)}_n (\tau, \hat \theta^{(b)}) = \bar \psi^{*(b)}_n (\tau, \hat \theta) + \frac{\partial \bar \psi^{*(b)}_n (\tau, \hat \theta)}{\partial \theta^{\intercal}} (\hat \theta^{(b)} - \hat \theta) + \frac{1}{2} (\hat \theta^{(b)} - \hat \theta)^{\intercal} \big\{\frac{\partial^2 \psi^{*(b)}_n (\tau, \Tilde{\theta}^{*(b)})}{\partial \theta \partial \theta^{\intercal}}\big\} (\hat \theta^{(b)} - \hat \theta).
\end{equation*}

By C5, $\partial^2 \psi^{*(b)}_n (\tau, \Tilde{\theta}^{*(b)})/(\partial \theta \partial \theta^{\intercal}) = O_p(1)$, thus the second term of the above equation is $o_p(n^{-1/2})$. Plug in \eqref{boot_thetab}, apply the same technique we use in the proof of Theorem 2, one can rewrite the above Taylor expansion as
\begin{align*}
    \bar \psi^{*(b)}_n (\tau, \hat \theta^{(b)}) = \frac{1}{n}\sum_{i = 1}^n u_{i}^{(b)} \big\{\sum_{m=1}^M w^{(m)}_i(\hat \theta^{(b)})\psi(Z_{i}^{*(m)}, \tau) + \Gamma^{(b)}(\tau, \hat \theta)^{\intercal}\bar s_i(\hat \theta) \big\} + o_p(n^{-1/2}),
\end{align*}
where $\Gamma^{(b)}(\tau, \hat \theta) = \mathbf{I}^{(b)-1}_{\text{obs}}(\hat \theta)\mathbf{I}^{(b)}_{\psi, \text{mis}}(\tau, \hat \theta)$. Here $\mathbf{I}^{(b)-1}_{\text{obs}}(\hat \theta) = n^{-1}\sum_{i=1}^n u^{(b)}_i \big\{\partial \bar s_i(\hat\theta)/\partial \theta^{\intercal}\big\}$ and
$$\mathbf{I}^{(b)}_{\psi, \text{mis}}(\tau, \hat \theta) = n^{-1}\sum_{i=1}^n u^{(b)}_i \sum_{m=1}^M w^{(m)}_i(\hat \theta)\big\{s(Z_{i}^{*(m)}, \hat \theta) - \bar s^*_i(\hat \theta)\big\}\psi(Z_{i}^{*(m)}, \tau).$$

Given both the observed and imputed data, by weak law of large number, $\mathbf{I}^{(b)}_{\text{obs}}(\hat \theta) \xrightarrow{\mathbb{P}} n^{-1}\sum_{i=1}^n \bar s^{*}_i(\hat \theta)   \bar s^{*}_i(\hat \theta)^{\intercal}$ and $$\mathbf{I}^{(b)}_{\psi, \text{mis}}(\tau, \hat \theta) \xrightarrow{\mathbb{P}} \Big[n^{-1}\sum_{i=1}^n\sum_{m=1}^M w^{(m)}_{i}(\hat \theta)\big\{s(Z^{*(m)}_{i}, \hat\theta) - \bar s^{*}_i(\hat \theta)\big\}\psi(Z^{*(m)}_{i}, \tau)\Big],$$ then we have $\Gamma^{(b)}(\tau, \hat \theta) - \hat \Gamma(\tau, \hat \theta) = o_p(1)$. Also by C6, $n^{-1}\sum_{i=1}^n u^{(b)}_i \bar s_i(\hat \theta) = O_p(n^{-1/2})$, then for any $\tau$, 

\begin{equation*}
    \bar \psi^{*(b)}_n (\tau, \hat \theta^{(b)}) = \frac{1}{n}\sum_{i = 1}^n u_{i}^{(b)} \big\{\sum_{m=1}^M w^{(m)}_i(\hat \theta)\psi(Z_{i}^{*(m)}, \tau) + \hat \Gamma(\tau, \hat \theta)^{\intercal}\bar s_i(\hat \theta) \big\} + o_p(n^{-1/2}).
\end{equation*}

Then follow the proof for asymptotic normality for M-estimation in Theorem 2, the asymptotic distribution of $\hat \tau^{(b)}_{DI,n}$ given the complete data is 
\begin{equation*}
    \sqrt n (\hat \tau^{(b)}_{DI,n} - \hat \tau_{DI,n}) \xrightarrow{d} \mathcal{N}(0, A^{(b)}(\hat \tau_{DI,n}, \hat \theta)^{-1}B^{(b)}(\hat \tau_{DI,n}, \hat \theta)\{A^{(b)}(\hat \tau_{DI,n}, \hat \theta)^{-1}\}^{\intercal}),
\end{equation*}
where 
\begin{align*}
    A^{(b)}(\hat \tau, \hat \theta) & = \lim_{n \rightarrow \infty} \frac{1}{n}\sum_{i=1}^n \mathbb{E}\Big[\frac{\partial \big\{u_{i}^{(b)}\sum_{m=1}^M w_i^{(m)}(\hat \theta)\psi (Z^{*(m)}_{i}, \hat \tau) + u_{i}^{(b)}\hat \Gamma(\hat \tau, \hat \theta)^{\intercal}\bar s_i(\hat \theta) \big\}}{\partial \tau} \big| Z \Big] \\
    & = \lim_{n \rightarrow \infty} \frac{1}{n}\sum_{i=1}^n \sum_{m=1}^M w_i^{(m)}(\hat \theta) \frac{\partial \psi (Z^{*(m)}_{i}, \hat \tau)} {\partial \tau} \\
    & = \lim_{n \rightarrow \infty} A_n(Z,\hat \tau, \hat \theta).
\end{align*}
Note that the second equation holds since $\sum_{i=1}^n \bar s_i(\hat \theta) = 0$ by \eqref{theta_score_eq}, and 
\begin{align*}
    B^{(b)}(\hat \tau, \hat \theta) & = \lim_{n \rightarrow \infty} \frac{1}{n}\sum_{i=1}^n \mathbb{V}\Big[u_{i}^{(b)}\big\{\sum_{m=1}^M w_i^{(m)}(\hat \theta)\psi (Z^{*(m)}_{i}, \hat \tau) + \hat \Gamma(\tau, \hat \theta)^{\intercal}\bar s_i(\hat \theta) \big\} \big| Z \Big] \\
    & = \lim_{n \rightarrow \infty} \frac{1}{n}\sum_{i=1}^n \big\{\sum_{m=1}^M w_i^{(m)}(\hat \theta)\psi (Z^{*(m)}_{i}, \hat \tau)\\ &\quad  + \hat \Gamma(\hat \tau, \hat \theta)^{\intercal}\bar s_i(\hat \theta) \big\} \big\{\sum_{m=1}^M w_i^{(m)}(\hat \theta)\psi (Z^{*(m)}_{i}, \hat \tau) + \hat \Gamma(\hat \tau, \hat \theta)^{\intercal}\bar s_i(\hat \theta) \big\}^{\intercal} \\
    & = \lim_{n \rightarrow \infty}B_n(Z,\hat \tau, \hat \theta).
\end{align*}

Note that by construction, $\hat{\mathbb{V}}_2(\hat \tau_{DI,n})$ is a consistent estimator of $$A^{(b)}(\hat \tau_{DI,n}, \hat \theta)^{-1}\big\{\frac{1}{n} B^{(b)}(\hat \tau_{DI,n}, \hat \theta)\big\}\{A^{(b)}(\hat \tau_{DI,n}, \hat \theta)^{-1}\}^{\intercal},$$ $A^{(b)}(\hat \tau, \hat \theta), B^{(b)}(\hat \tau, \hat \theta)$ are equivalent to $A_n(Z,\hat \tau, \hat \theta) , B_n(Z,\hat \tau, \hat \theta)$ when sample size $n$ is large, and $\allowbreak A_n(Z,\hat \tau, \hat \theta) ,  B_n(Z,\hat \tau, \hat \theta)$ are consistent estimators of $A(\tau_0, \theta_0), B(\tau_0, \theta_0)$. Therefore, $\hat{\mathbb{V}}_2(\hat \tau_{DI,n})$ is a consistent estimators of $\mathbb{V}(\hat \tau_{DI,n})$.

\end{proof}

\section{Additional simulation results}\label{appen:sim_result}

We generate the longitudinal responses in a sequential manner. The
baseline responses are generated by $Y_{i1}\mid(G_{i}=j)=\alpha_{j0}^{(1)}+X_{i}^{\intercal}\alpha_{j1}^{(1)}+\varepsilon_{ij1},$
where $\alpha_{j}^{(1)}=(\alpha_{j0}^{(1)},\alpha_{j1}^{(1)\intercal})^{\intercal}$
is set to be $\alpha_{0}^{(1)}=\alpha_{1}^{(1)}=(0.5,1,-3,2)^{\intercal}$
for both groups to mimic a randomized clinical trial. At $k$th visit
for $k=2,\cdots,5$, the sequential responses are generated by $Y_{ik}\mid(G_{i}=j)=\alpha_{j0}^{(k)}+X_{i}^{\intercal}\alpha_{j1}^{(k)}+\alpha_{j2}^{(k)}Y_{i1}+\cdots+\alpha_{jk}^{(k)}Y_{ik-1}+\varepsilon_{ijk}.$
For control group, set $\alpha_{10}^{(k)}\sim\mathcal{N}(\mu_{1k},\eta_{k}^{2})$
where $\mu_{1}:=(\mu_{12},\cdots,\mu_{15})^{\intercal}=(1,1.5,2,3)^{\intercal}$
and $\eta:=(\eta_{2},\cdots,\eta_{5})^{\intercal}=(0.5,1,1,1)^{\intercal}$,
$\alpha_{1l}^{(k)}\sim\mathcal{N}(0,0.5)$ for $l=1,\cdots,k-1$ and
$\alpha_{1k}^{(k)}\sim\text{Unif}(0,1)$ indicates a positive correlation
with the adjacent response. For treatment group, generate the regression
parameters as $\alpha_{20}^{(k)}\sim\mathcal{N}(\mu_{2k},\eta_{k}^{2})$
where $\mu_{2}:=(\mu_{22},\cdots,\mu_{25})^{\intercal}=(1.5,3,4.5,6)^{\intercal}$
represents different longitudinal responses for distinct groups, $\alpha_{2l}^{(k)}\sim\mathcal{N}(0,0.5)$
for $l=1,\cdots,k-1$ and $\alpha_{2k}^{(k)}\sim\text{Unif}(0,1)$.
The error terms are independently generated by $\varepsilon_{ijl}\sim\mathcal{N}(0,\sigma_{l}^{2})$
for $l=1,\cdots,T$, where $\sigma:=(\sigma_{1},\cdots,\sigma_{5})^{\intercal}=(2.0,1.8,2.0,2.1,2.2)^{\intercal}$
imitates an increase in variations for longitudinal outcomes in each
group.

Note that generating the longitudinal responses using the sequential
regression approach is equivalent to generating them from the multivariate
normal distribution following \eqref{eq:general_model}. Transform
the above sequential generating process, for the control and treatment
group, the coefficients $\beta_{j1},\cdots,\beta_{j5}$ used in the
simulation study are 
\begin{align*}
\begin{pmatrix}\beta_{11}^{\intercal}\\
\beta_{12}^{\intercal}\\
\beta_{13}^{\intercal}\\
\beta_{14}^{\intercal}\\
\beta_{15}^{\intercal}
\end{pmatrix}=\begin{pmatrix}0.50 & 1.00 & -3.00 & 2.00\\
0.73 & 0.80 & -1.46 & 0.16\\
1.55 & -0.07 & 1.31 & -0.09\\
2.19 & -0.08 & -1.35 & 0.95\\
4.29 & 0.62 & -1.76 & 1.30
\end{pmatrix};\begin{pmatrix}\beta_{21}^{\intercal}\\
\beta_{22}^{\intercal}\\
\beta_{23}^{\intercal}\\
\beta_{24}^{\intercal}\\
\beta_{25}^{\intercal}
\end{pmatrix}=\begin{pmatrix}0.50 & 1.00 & -3.00 & 2.00\\
2.16 & 1.08 & -2.24 & 1.23\\
7.31 & 0.39 & -3.29 & 0.88\\
6.45 & 1.05 & -0.22 & 0.18\\
5.82 & 0.09 & 0.83 & -0.47
\end{pmatrix}.
\end{align*}

The group specific covariance matrices $\Sigma^{(j)}$ for $j=1,2$
are 
\begin{align*}
\Sigma^{(1)}=\begin{pmatrix}4.00 & 2.66 & -0.63 & 1.58 & 1.93\\
2.66 & 5.01 & 0.34 & 1.10 & 1.81\\
-0.63 & 0.34 & 4.27 & 0.98 & 0.42\\
1.58 & 1.10 & 0.98 & 5.41 & 3.09\\
1.93 & 1.81 & 0.42 & 3.09 & 6.99
\end{pmatrix};\Sigma^{(2)}=\begin{pmatrix}4.00 & 2.91 & 2.28 & 0.12 & 0.21\\
2.91 & 5.36 & 4.74 & 1.99 & 0.73\\
2.28 & 4.74 & 8.23 & 2.63 & -0.22\\
0.12 & 1.99 & 2.63 & 5.67 & 0.37\\
0.21 & 0.73 & -0.22 & 0.37 & 5.16
\end{pmatrix}.
\end{align*}

Tables \ref{rtb_mean_table} and \ref{washout_mean_table} show the
simulation results of an ATE estimator under RTB and washout imputation
respectively. Similar to Table \ref{rbi_mean_table}, point estimates
from MI and DI become closer to the true value with smaller Monte
Carlo variances as sample size increases which confirms consistency.
MI and DI estimators have similar point estimates and Monte Carlo
variances under each imputation setting. Again, conservative variance
estimates take place under Rubin's method. The overestimation issue,
however, is relatively moderate compared to J2R setting. But one can
still observe relatively larger variances estimates compared to true
variances. Variance estimates using weighted bootstrap in DI outperforms
by smaller relative bias and a more precise coverage rate.

\begin{table}[!htbp]
\centering \caption{Simulation results under RTB of the ATE estimator. Here the true value
$\tau=1.5896$.}
\resizebox{\textwidth}{!}{%
\begin{tabular}{ccccccccccccccccccc}
\hline 
 &  & \multicolumn{2}{c}{Point est} &  & \multicolumn{2}{c}{True var} &  & \multicolumn{2}{c}{Var est} &  & \multicolumn{2}{c}{Relative bias} &  & \multicolumn{2}{c}{Coverage rate} &  & \multicolumn{2}{c}{Mean CI length}\tabularnewline
 &  & \multicolumn{2}{c}{($\times10^{-2}$)} &  & \multicolumn{2}{c}{($\times10^{-2}$)} &  & \multicolumn{2}{c}{($\times10^{-2}$)} &  & \multicolumn{2}{c}{($\%$)} &  & \multicolumn{2}{c}{($\%$)} &  & \multicolumn{2}{c}{($\times10^{-2}$)}\tabularnewline
\cline{3-4} \cline{4-4} \cline{6-7} \cline{7-7} \cline{9-10} \cline{10-10} \cline{12-13} \cline{13-13} \cline{15-16} \cline{16-16} \cline{18-19} \cline{19-19} 
N & M & MI & DI &  & MI & DI &  & MI & DI &  & MI & DI &  & MI & DI &  & MI & DI\tabularnewline
\hline 
 & 5 & 156.32 & 155.70 &  & 22.48 & 22.36 &  & 25.88 & 21.82 &  & 15.14 & -2.41 &  & 96.20 & 94.60 &  & 199.12 & 182.34\tabularnewline
100 & 10 & 156.13 & 155.70 &  & 22.20 & 22.30 &  & 25.64 & 21.77 &  & 15.47 & -2.41 &  & 96.30 & 94.60 &  & 198.22 & 182.12\tabularnewline
 & 100 & 156.08 & 156.02 &  & 22.17 & 22.09 &  & 25.55 & 21.72 &  & 15.26 & -1.68 &  & 96.20 & 94.90 &  & 197.91 & 181.92\tabularnewline
\hline 
 & 5 & 159.29 & 159.19 &  & 4.38 & 4.45 &  & 5.04 & 4.51 &  & 15.25 & 1.22 &  & 97.10 & 94.70 &  & 87.99 & 82.97\tabularnewline
500 & 10 & 159.32 & 159.32 &  & 4.36 & 4.41 &  & 5.04 & 4.50 &  & 15.63 & 2.09 &  & 97.30 & 95.10 &  & 87.96 & 82.89\tabularnewline
 & 100 & 159.28 & 159.35 &  & 4.36 & 4.35 &  & 5.00 & 4.49 &  & 14.65 & 3.14 &  & 96.90 & 95.20 &  & 87.60 & 82.82\tabularnewline
\hline 
 & 5 & 159.19 & 159.11 &  & 2.13 & 2.15 &  & 2.52 & 2.26 &  & 18.71 & 5.05 &  & 96.10 & 95.10 &  & 62.26 & 58.81\tabularnewline
1000 & 10 & 159.10 & 159.11 &  & 2.15 & 2.12 &  & 2.51 & 2.26 &  & 16.88 & 6.62 &  & 96.10 & 95.30 &  & 62.07 & 58.76\tabularnewline
 & 100 & 159.18 & 159.14 &  & 2.12 & 2.10 &  & 2.50 & 2.25 &  & 17.51 & 7.16 &  & 96.50 & 95.00 &  & 61.91 & 58.72\tabularnewline
\hline 
\end{tabular}} \label{rtb_mean_table}
\end{table}

\begin{table}[!htbp]
\centering \caption{Simulation results under J2R of the ATE estimator. Here the true value
$\tau=0.7858$.}
\resizebox{\textwidth}{!}{%
\begin{tabular}{ccccccccccccccccccc}
\hline 
 &  & \multicolumn{2}{c}{Point est} &  & \multicolumn{2}{c}{True var} &  & \multicolumn{2}{c}{Var est} &  & \multicolumn{2}{c}{Relative bias} &  & \multicolumn{2}{c}{Coverage rate} &  & \multicolumn{2}{c}{Mean CI length}\tabularnewline
 &  & \multicolumn{2}{c}{($\times10^{-2}$)} &  & \multicolumn{2}{c}{($\times10^{-2}$)} &  & \multicolumn{2}{c}{($\times10^{-2}$)} &  & \multicolumn{2}{c}{($\%$)} &  & \multicolumn{2}{c}{($\%$)} &  & \multicolumn{2}{c}{($\times10^{-2}$)}\tabularnewline
\cline{3-4} \cline{4-4} \cline{6-7} \cline{7-7} \cline{9-10} \cline{10-10} \cline{12-13} \cline{13-13} \cline{15-16} \cline{16-16} \cline{18-19} \cline{19-19} 
N & M & MI & DI &  & MI & DI &  & MI & DI &  & MI & DI &  & MI & DI &  & MI & DI\tabularnewline
\hline 
 & 5 & 75.35 & 74.55 &  & 22.07 & 22.10 &  & 24.22 & 21.02 &  & 9.74 & -4.88 &  & 96.00 & 94.40 &  & 192.54 & 178.92\tabularnewline
100 & 10 & 75.05 & 74.63 &  & 21.75 & 21.96 &  & 23.96 & 21.18 &  & 10.17 & -3.56 &  & 95.80 & 94.10 &  & 191.59 & 179.61\tabularnewline
 & 100 & 75.01 & 74.94 &  & 21.70 & 21.62 &  & 23.84 & 21.33 &  & 9.87 & -1.38 &  & 96.30 & 94.90 &  & 191.14 & 180.23\tabularnewline
\hline 
 & 5 & 78.79 & 78.69 &  & 4.38 & 4.47 &  & 4.71 & 4.35 &  & 7.59 & -2.71 &  & 96.10 & 94.80 &  & 85.00 & 81.54\tabularnewline
500 & 10 & 78.82 & 78.83 &  & 4.37 & 4.39 &  & 4.70 & 4.38 &  & 7.61 & -0.43 &  & 96.20 & 95.30 &  & 84.94 & 81.76\tabularnewline
 & 100 & 78.79 & 78.87 &  & 4.34 & 4.35 &  & 4.65 & 4.40 &  & 7.07 & 1.28 &  & 96.60 & 95.60 &  & 84.49 & 82.00\tabularnewline
\hline 
 & 5 & 79.01 & 78.92 &  & 2.15 & 2.16 &  & 2.36 & 2.19 &  & 9.41 & 1.30 &  & 95.90 & 94.40 &  & 60.15 & 57.87\tabularnewline
1000 & 10 & 78.92 & 78.92 &  & 2.16 & 2.13 &  & 2.34 & 2.20 &  & 8.40 & 3.53 &  & 95.50 & 94.40 &  & 59.91 & 58.05\tabularnewline
 & 100 & 79.00 & 78.95 &  & 2.13 & 2.12 &  & 2.32 & 2.22 &  & 8.85 & 4.71 &  & 95.60 & 94.80 &  & 59.73 & 58.21\tabularnewline
\hline 
\end{tabular}} \label{washout_mean_table}
\end{table}

Tables \ref{rtb_prop_table}--\ref{washout_prop_table} present the
results estimating risk difference under RTB, J2R and washout imputation
respectively. The performances are similar to the previous cases.
Again, similar to the cases of estimation of a regression type of
ATE, MI with Rubin's variance estimator overestimates the true variance
since it has much larger variance estimates under each imputation
assumption. In most cases under RTB and washout imputation assumption,
it is moderately conservative compared to the one under J2R. DI with
variance estimates obtained from weighted bootstrap outperforms MI
with Rubin's estimates by the proximity to true variances, much smaller
relative bias, and more precise coverage probabilities in most cases
under each imputation assumption. The variance estimates are close
to true variances, with relative bias controlled under $5\%$, and
do not show a tendency of overestimation or underestimation. The coverage
probabilities are around $95\%$ in a tolerable range.

\begin{table}[!htbp]
\centering \caption{Simulation results under RTB of the risk difference estimator. Here
the true value $\tau=0.2192$.}
\resizebox{\textwidth}{!}{%
\begin{tabular}{ccccccccccccccccccc}
\hline 
 &  & \multicolumn{2}{c}{Point est} &  & \multicolumn{2}{c}{True var} &  & \multicolumn{2}{c}{Var est} &  & \multicolumn{2}{c}{Relative bias} &  & \multicolumn{2}{c}{Coverage rate} &  & \multicolumn{2}{c}{Mean CI length}\tabularnewline
 &  & \multicolumn{2}{c}{($\times10^{-2}$)} &  & \multicolumn{2}{c}{($\times10^{-4}$)} &  & \multicolumn{2}{c}{($\times10^{-4}$)} &  & \multicolumn{2}{c}{($\%$)} &  & \multicolumn{2}{c}{($\%$)} &  & \multicolumn{2}{c}{($\times10^{-2}$)}\tabularnewline
\cline{3-4} \cline{4-4} \cline{6-7} \cline{7-7} \cline{9-10} \cline{10-10} \cline{12-13} \cline{13-13} \cline{15-16} \cline{16-16} \cline{18-19} \cline{19-19} 
N & M & MI & DI &  & MI & DI &  & MI & DI &  & MI & DI &  & MI & DI &  & MI & DI\tabularnewline
\hline 
 & 5 & 21.71 & 21.71 &  & 42.75 & 42.48 &  & 51.37 & 44.37 &  & 20.14 & 4.45 &  & 97.20 & 95.30 &  & 28.08 & 26.04\tabularnewline
100 & 10 & 21.70 & 21.67 &  & 41.96 & 42.21 &  & 50.99 & 44.09 &  & 21.53 & 4.46 &  & 97.40 & 96.00 &  & 27.98 & 25.96\tabularnewline
 & 100 & 21.72 & 21.72 &  & 42.02 & 42.00 &  & 50.72 & 43.86 &  & 20.71 & 4.44 &  & 97.30 & 96.00 &  & 27.91 & 25.89\tabularnewline
\hline 
 & 5 & 21.93 & 21.89 &  & 9.48 & 9.64 &  & 10.32 & 9.10 &  & 8.91 & -5.62 &  & 96.00 & 93.70 &  & 12.59 & 11.80\tabularnewline
500 & 10 & 21.91 & 21.91 &  & 9.35 & 9.45 &  & 10.28 & 9.05 &  & 9.93 & -4.28 &  & 95.70 & 94.40 &  & 12.56 & 11.76\tabularnewline
 & 100 & 21.91 & 21.91 &  & 9.35 & 9.32 &  & 10.21 & 9.01 &  & 9.26 & -3.39 &  & 95.70 & 94.70 &  & 12.53 & 11.73\tabularnewline
\hline 
 & 5 & 21.93 & 21.93 &  & 4.44 & 4.40 &  & 5.17 & 4.55 &  & 16.51 & 3.40 &  & 96.70 & 95.90 &  & 8.91 & 8.35\tabularnewline
1000 & 10 & 21.93 & 21.93 &  & 4.35 & 4.32 &  & 5.15 & 4.53 &  & 18.35 & 4.74 &  & 97.10 & 95.50 &  & 8.89 & 8.32\tabularnewline
 & 100 & 21.94 & 21.93 &  & 4.36 & 4.33 &  & 5.11 & 4.51 &  & 17.34 & 4.14 &  & 97.60 & 95.70 &  & 8.86 & 8.31\tabularnewline
\hline 
\end{tabular}} \label{rtb_prop_table}
\end{table}

\begin{table}[!htbp]
\centering \caption{Simulation results under J2R of the risk difference estimator. Here
the true value $\tau=0.2197$.}
\resizebox{\textwidth}{!}{%
\begin{tabular}{ccccccccccccccccccc}
\hline 
 &  & \multicolumn{2}{c}{Point est} &  & \multicolumn{2}{c}{True var} &  & \multicolumn{2}{c}{Var est} &  & \multicolumn{2}{c}{Relative bias} &  & \multicolumn{2}{c}{Coverage rate} &  & \multicolumn{2}{c}{Mean CI length}\tabularnewline
 &  & \multicolumn{2}{c}{($\times10^{-2}$)} &  & \multicolumn{2}{c}{($\times10^{-4}$)} &  & \multicolumn{2}{c}{($\times10^{-4}$)} &  & \multicolumn{2}{c}{($\%$)} &  & \multicolumn{2}{c}{($\%$)} &  & \multicolumn{2}{c}{($\times10^{-2}$)}\tabularnewline
\cline{3-4} \cline{4-4} \cline{6-7} \cline{7-7} \cline{9-10} \cline{10-10} \cline{12-13} \cline{13-13} \cline{15-16} \cline{16-16} \cline{18-19} \cline{19-19} 
N & M & MI & DI &  & MI & DI &  & MI & DI &  & MI & DI &  & MI & DI &  & MI & DI\tabularnewline
\hline 
 & 5 & 21.65 & 21.64 &  & 39.02 & 38.53 &  & 53.60 & 39.94 &  & 37.37 & 3.67 &  & 98.00 & 95.30 &  & 28.66 & 24.71\tabularnewline
100 & 10 & 21.69 & 21.61 &  & 37.07 & 37.75 &  & 52.95 & 39.31 &  & 42.84 & 4.15 &  & 98.30 & 95.30 &  & 28.50 & 24.51\tabularnewline
 & 100 & 21.64 & 21.64 &  & 37.24 & 37.08 &  & 52.18 & 38.67 &  & 40.14 & 4.28 &  & 98.10 & 95.10 &  & 28.31 & 24.31\tabularnewline
\hline 
 & 5 & 21.86 & 21.81 &  & 8.34 & 8.32 &  & 10.79 & 8.15 &  & 29.35 & -2.04 &  & 96.60 & 94.80 &  & 12.86 & 11.16\tabularnewline
500 & 10 & 21.85 & 21.85 &  & 8.23 & 8.32 &  & 10.61 & 8.01 &  & 28.90 & -3.72 &  & 97.30 & 94.70 &  & 12.76 & 11.06\tabularnewline
 & 100 & 21.84 & 21.85 &  & 8.16 & 8.15 &  & 10.49 & 7.89 &  & 28.54 & -3.23 &  & 97.30 & 94.50 &  & 12.70 & 10.98\tabularnewline
\hline 
 & 5 & 21.98 & 21.95 &  & 4.22 & 4.20 &  & 5.38 & 4.07 &  & 27.35 & -2.94 &  & 97.30 & 94.60 &  & 9.08 & 7.89\tabularnewline
1000 & 10 & 21.97 & 21.96 &  & 4.15 & 4.09 &  & 5.30 & 4.00 &  & 27.67 & -2.18 &  & 97.80 & 94.80 &  & 9.02 & 7.82\tabularnewline
 & 100 & 21.96 & 21.97 &  & 4.06 & 4.04 &  & 5.25 & 3.94 &  & 29.29 & -2.57 &  & 97.70 & 94.80 &  & 8.98 & 7.76\tabularnewline
\hline 
\end{tabular}} \label{rbi_prop_table}
\end{table}

\begin{table}[!htbp]
\centering \caption{Simulation results under washout of the risk difference estimator.
Here the true value $\tau=0.1478$.}
\resizebox{\textwidth}{!}{%
\begin{tabular}{ccccccccccccccccccc}
\hline 
 &  & \multicolumn{2}{c}{Point est} &  & \multicolumn{2}{c}{True var} &  & \multicolumn{2}{c}{Var est} &  & \multicolumn{2}{c}{Relative bias} &  & \multicolumn{2}{c}{Coverage rate} &  & \multicolumn{2}{c}{Mean CI length}\tabularnewline
 &  & \multicolumn{2}{c}{($\times10^{-2}$)} &  & \multicolumn{2}{c}{($\times10^{-4}$)} &  & \multicolumn{2}{c}{($\times10^{-4}$)} &  & \multicolumn{2}{c}{($\%$)} &  & \multicolumn{2}{c}{($\%$)} &  & \multicolumn{2}{c}{($\times10^{-2}$)}\tabularnewline
\cline{3-4} \cline{4-4} \cline{6-7} \cline{7-7} \cline{9-10} \cline{10-10} \cline{12-13} \cline{13-13} \cline{15-16} \cline{16-16} \cline{18-19} \cline{19-19} 
N & M & MI & DI &  & MI & DI &  & MI & DI &  & MI & DI &  & MI & DI &  & MI & DI\tabularnewline
\hline 
 & 5 & 14.58 & 14.56 &  & 44.97 & 44.85 &  & 54.15 & 46.13 &  & 20.41 & 2.87 &  & 97.30 & 94.80 &  & 28.82 & 26.56\tabularnewline
100 & 10 & 14.56 & 14.51 &  & 44.06 & 44.70 &  & 53.48 & 46.03 &  & 21.37 & 2.99 &  & 96.80 & 95.00 &  & 28.66 & 26.52\tabularnewline
 & 100 & 14.57 & 14.57 &  & 44.16 & 44.05 &  & 53.08 & 45.93 &  & 20.21 & 4.27 &  & 96.60 & 95.10 &  & 28.56 & 26.50\tabularnewline
\hline 
 & 5 & 14.77 & 14.78 &  & 9.87 & 10.06 &  & 10.87 & 9.50 &  & 10.14 & -5.64 &  & 95.80 & 94.00 &  & 12.91 & 12.05\tabularnewline
500 & 10 & 14.76 & 14.78 &  & 9.74 & 9.88 &  & 10.80 & 9.46 &  & 10.92 & -4.21 &  & 95.30 & 94.30 &  & 12.88 & 12.03\tabularnewline
 & 100 & 14.77 & 14.78 &  & 9.75 & 9.71 &  & 10.70 & 9.45 &  & 9.79 & -2.63 &  & 95.70 & 94.50 &  & 12.82 & 12.02\tabularnewline
\hline 
 & 5 & 14.80 & 14.81 &  & 4.82 & 4.78 &  & 5.43 & 4.76 &  & 12.53 & -0.36 &  & 96.50 & 94.20 &  & 9.13 & 8.54\tabularnewline
1000 & 10 & 14.80 & 14.80 &  & 4.77 & 4.71 &  & 5.41 & 4.75 &  & 13.41 & 1.01 &  & 96.30 & 94.40 &  & 9.11 & 8.53\tabularnewline
 & 100 & 14.81 & 14.80 &  & 4.73 & 4.71 &  & 5.35 & 4.74 &  & 13.24 & 0.79 &  & 96.40 & 94.70 &  & 9.07 & 8.52\tabularnewline
\hline 
\end{tabular}} \label{washout_prop_table}
\end{table}

Tables \ref{rtb_quantile_table}--\ref{washout_quantile_table} show
the results estimating QTE under RTB and washout imputation respectively.
MI and DI estimators have similar point estimates and Monte Carlo
variances under each imputation setting. In terms of variance estimates,
unlike the ones regarding ATE and risk difference, when the sample
size is relatively small, variance estimates of QTE from both methods
overestimate the true variance with large relative biases. The variance
estimates using Rubin's method for MI estimator are much larger than
true variances, resulting in large relative bias and very conservative
coverage rates. Under RTB and washout imputation, the overestimation
issue is less severe compared to the one under J2R as the sample size
increases. However, variance estimates from DI overestimate the true
variance when sample size $N=100$. With a small sample size, the
relative bias for DI estimator appears to be unsatisfying, while the
coverage rate does not show much overestimation. It may be due to
the instability of point estimates since the quantile estimator may
be skewed. Variance estimates become closer to true values with small
relative bias as the sample size grows. The majority of coverage rates
are close to the empirical values except for Table \ref{rbi_quantile_table}
when $N=500$ and $M=5$, which may be due to some Monte Carlo error.
Large sample size is recommended when estimating QTE based on the
simulation results.

\begin{table}[!htbp]
\centering \caption{Simulation results under RTB of the QTE estimator. Here the true value
of $\tau=1.8120$.}
\resizebox{\textwidth}{!}{%
\begin{tabular}{ccccccccccccccccccc}
\hline 
 &  & \multicolumn{2}{c}{Point est} &  & \multicolumn{2}{c}{True var} &  & \multicolumn{2}{c}{Var est} &  & \multicolumn{2}{c}{Relative bias} &  & \multicolumn{2}{c}{Coverage rate} &  & \multicolumn{2}{c}{Mean CI length}\tabularnewline
 &  & \multicolumn{2}{c}{($\times10^{-2}$)} &  & \multicolumn{2}{c}{($\times10^{-2}$)} &  & \multicolumn{2}{c}{($\times10^{-2}$)} &  & \multicolumn{2}{c}{($\%$)} &  & \multicolumn{2}{c}{($\%$)} &  & \multicolumn{2}{c}{($\times10^{-2}$)}\tabularnewline
\cline{3-4} \cline{4-4} \cline{6-7} \cline{7-7} \cline{9-10} \cline{10-10} \cline{12-13} \cline{13-13} \cline{15-16} \cline{16-16} \cline{18-19} \cline{19-19} 
N & M & MI & DI &  & MI & DI &  & MI & DI &  & MI & DI &  & MI & DI &  & MI & DI\tabularnewline
\hline 
 & 5 & 180.30 & 180.75 &  & 30.58 & 31.11 &  & 42.45 & 35.38 &  & 38.83 & 13.73 &  & 97.60 & 94.40 &  & 254.19 & 229.74\tabularnewline
100 & 10 & 180.53 & 180.32 &  & 30.04 & 30.86 &  & 42.26 & 35.42 &  & 40.70 & 14.79 &  & 97.70 & 95.10 &  & 253.71 & 229.83\tabularnewline
 & 100 & 180.48 & 180.53 &  & 29.84 & 30.80 &  & 42.04 & 35.39 &  & 40.85 & 14.90 &  & 97.80 & 94.90 &  & 253.07 & 229.76\tabularnewline
\hline 
 & 5 & 181.46 & 181.14 &  & 6.69 & 6.87 &  & 8.05 & 6.90 &  & 20.37 & 0.43 &  & 96.30 & 94.70 &  & 111.07 & 102.12\tabularnewline
500 & 10 & 181.27 & 181.37 &  & 6.58 & 6.67 &  & 7.99 & 6.87 &  & 21.53 & 3.10 &  & 96.20 & 94.30 &  & 110.68 & 101.94\tabularnewline
 & 100 & 181.31 & 181.30 &  & 6.56 & 6.64 &  & 7.95 & 6.84 &  & 21.22 & 2.98 &  & 96.30 & 94.50 &  & 110.40 & 101.71\tabularnewline
\hline 
 & 5 & 181.30 & 181.26 &  & 3.29 & 3.37 &  & 3.93 & 3.40 &  & 19.36 & 0.87 &  & 96.80 & 94.20 &  & 77.63 & 71.82\tabularnewline
1000 & 10 & 181.24 & 181.29 &  & 3.29 & 3.34 &  & 3.92 & 3.38 &  & 19.11 & 1.32 &  & 97.10 & 94.00 &  & 77.57 & 71.61\tabularnewline
 & 100 & 181.32 & 181.27 &  & 3.28 & 3.30 &  & 3.90 & 3.37 &  & 18.88 & 2.20 &  & 96.90 & 94.20 &  & 77.33 & 71.50\tabularnewline
\hline 
\end{tabular}} \label{rtb_quantile_table}
\end{table}

\begin{table}[!htbp]
\centering \caption{Simulation results under J2R of the QTE estimator. Here the true value
$\tau=1.5570$.}
\resizebox{\textwidth}{!}{%
\begin{tabular}{ccccccccccccccccccc}
\hline 
 &  & \multicolumn{2}{c}{Point est} &  & \multicolumn{2}{c}{True var} &  & \multicolumn{2}{c}{Var est} &  & \multicolumn{2}{c}{Relative bias} &  & \multicolumn{2}{c}{Coverage rate} &  & \multicolumn{2}{c}{Mean CI length}\tabularnewline
 &  & \multicolumn{2}{c}{($\times10^{-2}$)} &  & \multicolumn{2}{c}{($\times10^{-2}$)} &  & \multicolumn{2}{c}{($\times10^{-2}$)} &  & \multicolumn{2}{c}{($\%$)} &  & \multicolumn{2}{c}{($\%$)} &  & \multicolumn{2}{c}{($\times10^{-2}$)}\tabularnewline
\cline{3-4} \cline{4-4} \cline{6-7} \cline{7-7} \cline{9-10} \cline{10-10} \cline{12-13} \cline{13-13} \cline{15-16} \cline{16-16} \cline{18-19} \cline{19-19} 
N & M & MI & DI &  & MI & DI &  & MI & DI &  & MI & DI &  & MI & DI &  & MI & DI\tabularnewline
\hline 
 & 5 & 153.43 & 153.00 &  & 24.95 & 25.67 &  & 39.31 & 29.32 &  & 57.58 & 14.21 &  & 98.60 & 95.10 &  & 244.22 & 208.96\tabularnewline
100 & 10 & 153.47 & 152.83 &  & 24.09 & 25.13 &  & 38.57 & 29.10 &  & 60.07 & 15.80 &  & 98.30 & 95.20 &  & 242.12 & 208.12\tabularnewline
 & 100 & 153.22 & 153.17 &  & 23.72 & 24.41 &  & 38.15 & 28.87 &  & 60.87 & 18.31 &  & 98.40 & 95.20 &  & 241.00 & 207.39\tabularnewline
\hline 
 & 5 & 155.56 & 155.12 &  & 5.69 & 5.80 &  & 7.44 & 5.69 &  & 30.80 & -1.82 &  & 97.40 & 92.70 &  & 106.64 & 92.84\tabularnewline
500 & 10 & 155.25 & 155.38 &  & 5.66 & 5.76 &  & 7.32 & 5.60 &  & 29.32 & -2.62 &  & 97.10 & 94.10 &  & 105.88 & 92.13\tabularnewline
 & 100 & 155.27 & 155.41 &  & 5.59 & 5.67 &  & 7.24 & 5.55 &  & 29.53 & -2.11 &  & 97.30 & 93.20 &  & 105.36 & 91.68\tabularnewline
\hline 
 & 5 & 155.94 & 155.80 &  & 2.66 & 2.67 &  & 3.64 & 2.80 &  & 36.73 & 4.84 &  & 98.40 & 95.80 &  & 74.64 & 65.17\tabularnewline
1000 & 10 & 155.92 & 155.74 &  & 2.62 & 2.62 &  & 3.59 & 2.76 &  & 37.00 & 5.51 &  & 98.20 & 95.80 &  & 74.19 & 64.76\tabularnewline
 & 100 & 155.83 & 155.82 &  & 2.58 & 2.57 &  & 3.55 & 2.73 &  & 37.79 & 6.23 &  & 98.50 & 95.90 &  & 73.85 & 64.37\tabularnewline
\hline 
\end{tabular}} \label{rbi_quantile_table}
\end{table}

\begin{table}[!htbp]
\centering \caption{Simulation results under washout of the QTE estimator. Here the true
value $\tau=1.1313$.}
\resizebox{\textwidth}{!}{%
\begin{tabular}{ccccccccccccccccccc}
\hline 
 &  & \multicolumn{2}{c}{Point est} &  & \multicolumn{2}{c}{True var} &  & \multicolumn{2}{c}{Var est} &  & \multicolumn{2}{c}{Relative bias} &  & \multicolumn{2}{c}{Coverage rate} &  & \multicolumn{2}{c}{Mean CI length}\tabularnewline
 &  & \multicolumn{2}{c}{($\times10^{-2}$)} &  & \multicolumn{2}{c}{($\times10^{-2}$)} &  & \multicolumn{2}{c}{($\times10^{-2}$)} &  & \multicolumn{2}{c}{($\%$)} &  & \multicolumn{2}{c}{($\%$)} &  & \multicolumn{2}{c}{($\times10^{-2}$)}\tabularnewline
\cline{3-4} \cline{4-4} \cline{6-7} \cline{7-7} \cline{9-10} \cline{10-10} \cline{12-13} \cline{13-13} \cline{15-16} \cline{16-16} \cline{18-19} \cline{19-19} 
N & M & MI & DI &  & MI & DI &  & MI & DI &  & MI & DI &  & MI & DI &  & MI & DI\tabularnewline
\hline 
 & 5 & 111.26 & 111.43 &  & 28.85 & 29.72 &  & 40.59 & 33.64 &  & 40.69 & 13.20 &  & 97.30 & 94.80 &  & 248.45 & 223.75\tabularnewline
100 & 10 & 111.53 & 111.29 &  & 28.49 & 29.23 &  & 40.26 & 33.97 &  & 41.34 & 16.21 &  & 97.00 & 94.70 &  & 247.52 & 225.00\tabularnewline
 & 100 & 111.41 & 111.77 &  & 28.23 & 29.15 &  & 40.01 & 34.07 &  & 41.73 & 16.87 &  & 97.20 & 95.00 &  & 246.82 & 225.32\tabularnewline
\hline 
 & 5 & 113.38 & 113.42 &  & 6.60 & 6.82 &  & 7.70 & 6.62 &  & 16.56 & -3.01 &  & 96.80 & 93.70 &  & 108.53 & 100.08\tabularnewline
500 & 10 & 113.28 & 113.45 &  & 6.48 & 6.70 &  & 7.62 & 6.60 &  & 17.75 & -1.50 &  & 96.40 & 93.70 &  & 108.09 & 99.98\tabularnewline
 & 100 & 113.33 & 113.38 &  & 6.52 & 6.63 &  & 7.56 & 6.60 &  & 16.10 & -0.37 &  & 97.00 & 93.50 &  & 107.69 & 99.99\tabularnewline
\hline 
 & 5 & 113.43 & 113.37 &  & 3.19 & 3.20 &  & 3.76 & 3.28 &  & 17.88 & 2.67 &  & 97.40 & 95.20 &  & 75.88 & 70.55\tabularnewline
1000 & 10 & 113.47 & 113.46 &  & 3.18 & 3.15 &  & 3.75 & 3.28 &  & 18.03 & 3.98 &  & 97.00 & 95.20 &  & 75.85 & 70.53\tabularnewline
 & 100 & 113.48 & 113.45 &  & 3.13 & 3.13 &  & 3.71 & 3.28 &  & 18.51 & 4.78 &  & 97.20 & 95.30 &  & 75.44 & 70.51\tabularnewline
\hline 
\end{tabular}} \label{washout_quantile_table}
\end{table}

\section{Real data application}\label{appen:real_data}

\subsection{Additional analysis results}\label{appen:real_result}

The public dataset is available at \url{https://www.lshtm.ac.uk/research/centres-projects-groups/missing-data\#dia-missing-data}.
For the ATE, we estimate it by fitting a group-specific ANCOVA model
with the baseline HAMD-17 score as the baseline covariate. For the
risk difference, we are interested in the percentage difference of
patients with $50\%$ or more improvement from the baseline HAMD-17
score at the end of the trial between the control and treatment groups.
We estimate it through the estimating equations in Example \ref{example2}
(b). For the QTE, we do not limit on one specific quantile; instead,
we present the estimated cumulative distribution function (CDF) of
the relative change from baseline for each group obtained from the
estimating equations in Example \ref{example2} (c). In the implementation
of MI and DI, the imputation size $M=100$, and the number of bootstrap
replicates $B=100$. For all the hypothesis tests, we choose the significance
level $\alpha=0.05$.

\begin{figure}[!htbp]
\includegraphics[width=0.49\textwidth]{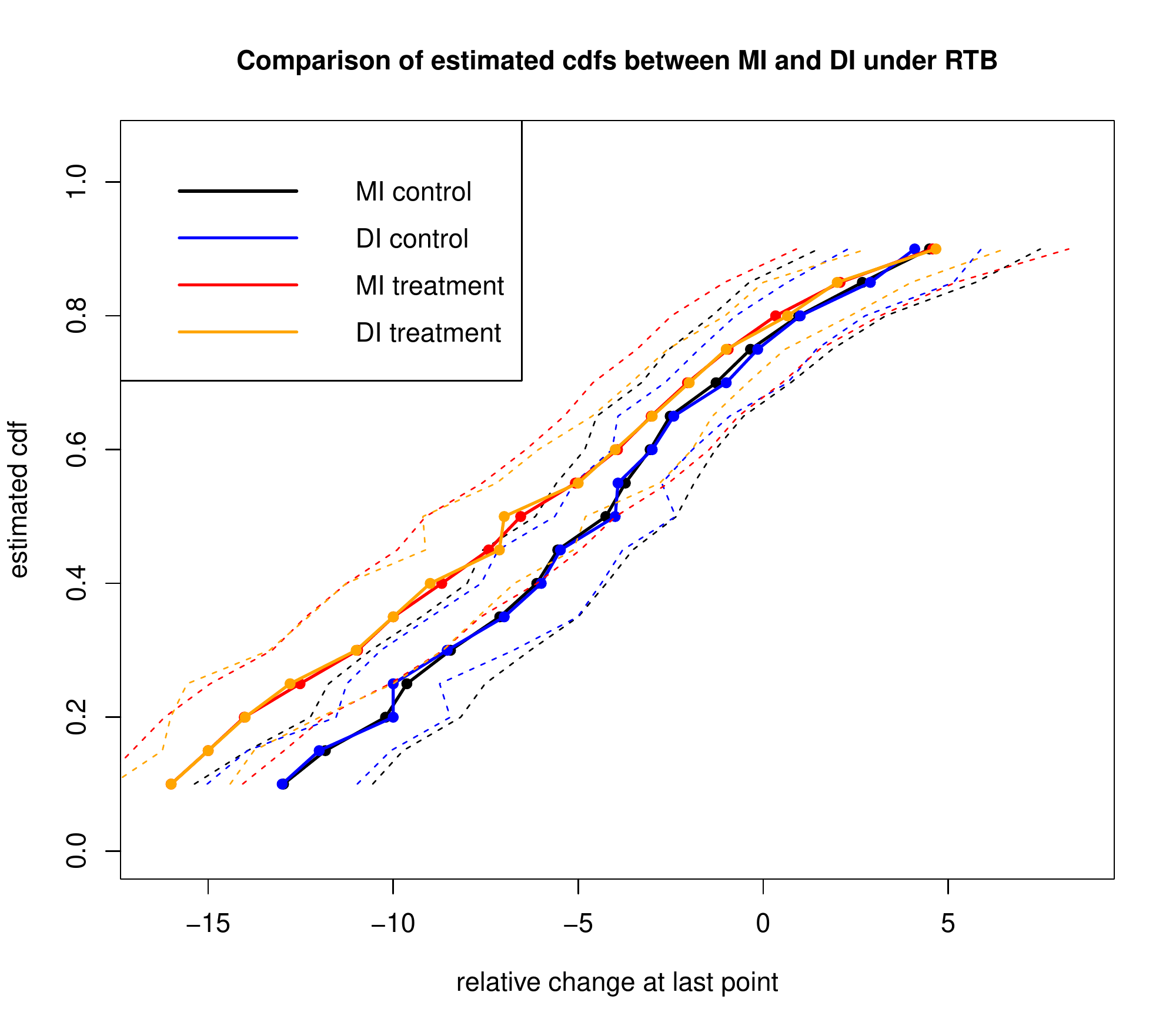}
\includegraphics[width=0.49\textwidth]{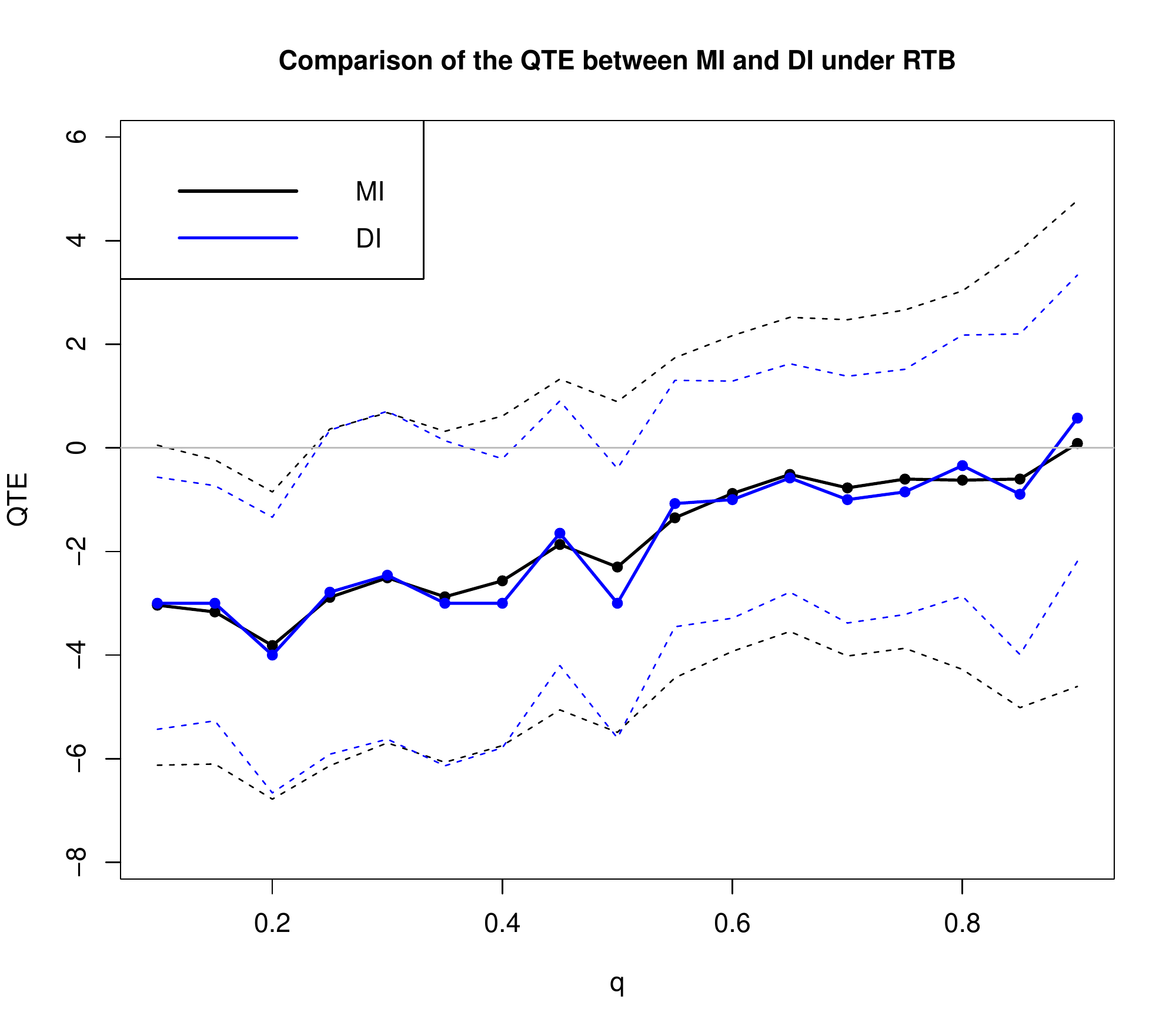}
\caption{The left figure is the estimated cdf of relative change from baseline at last time point via MI and DI in each group under RTB; the right figure is the estimated QTE as a function for a particular value of $q$. Both plots are accompanied by the point-wise $95\%$ CI in dashed lines.}
\label{fig:rtb}
\end{figure}

\begin{figure}[!htbp]
\includegraphics[width=0.49\textwidth]{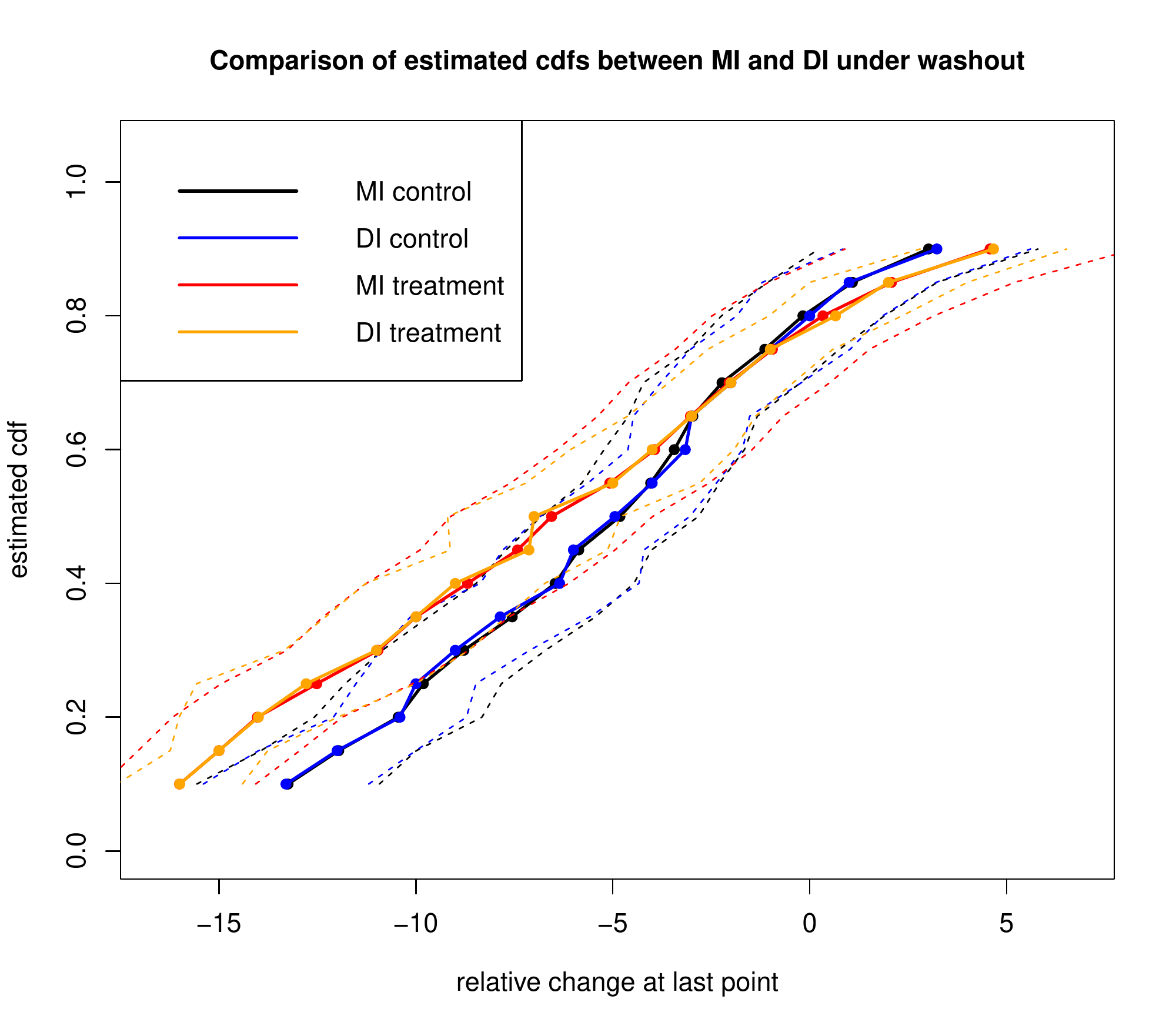}
\includegraphics[width=0.49\textwidth]{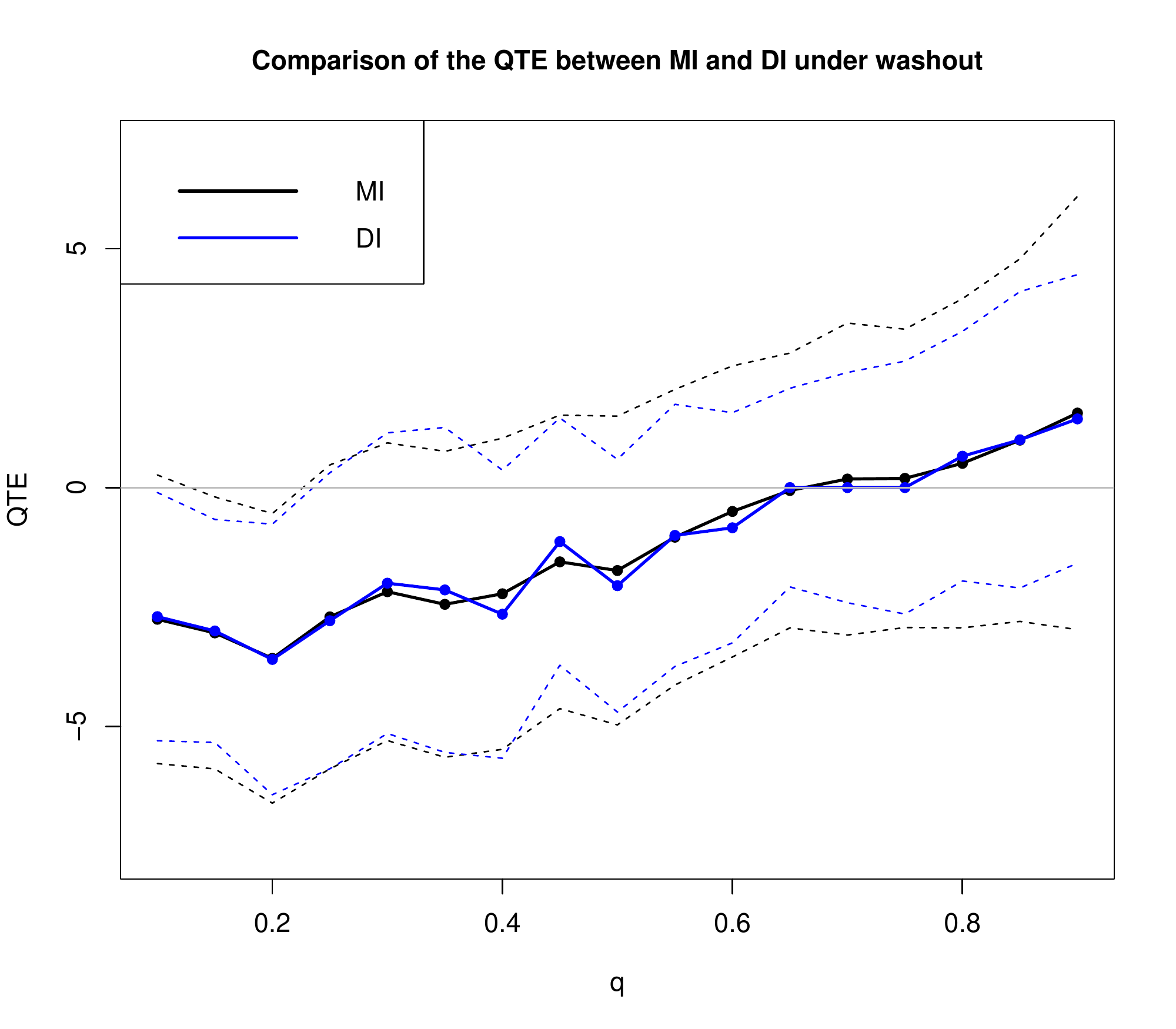}
\caption{The left figure is the estimated cdf of relative change from baseline at last time point via MI and DI in each group under washout imputation; the right figure is the estimated QTE as a function for a particular value of $q$. Both plots are accompanied by the point-wise $95\%$ CI in dashed lines.}
\label{fig:washout}
\end{figure}

We present the estimated CDFs of
the relative change at the last time point via MI and DI in each group
accompanied by the point-wise $95\%$ CI under RTB and washout imputation
mechanism in the left side of Figures \ref{fig:rtb} and \ref{fig:washout},
and the estimated QTEs accompanied by the point-wise $95\%$ CI as
a function of $q$ denoted as quantile percentage in the right side
of Figures \ref{fig:rtb} and \ref{fig:washout}. Similar to results
in the simulation study and the results under J2R, the estimated CDF
obtained from DI has a comparable shape as the one from MI with a
narrower $95\%$ confidence region under those two imputation mechanisms.
From the figure of the estimated QTE, a significant effect of the treatment
is detected for patients in the lower quantile of HAMD-17 score in
all three sensitivity analysis settings.

%\subsection{Compare weighted bootstrap with nonparametric bootstrap}\label{appen:real_nonpara}

%We compare the variance estimation results for the DI estimator using the proposed weighted bootstrap approach with the conventional nonparametric bootstrap method as presented in Tables \ref{tab:real_reg_nonpara} and \ref{tab:real_prop_nonpara}. We observe parallel performance for the two replication-based variance estimation methods. Both methods draw the same conclusion of the hypothesis test $H_0: \tau = 0$ regarding the treatment effect. Note that for the risk difference under RTB, we observe the p-values of the test as $0.049$ and $0.053$, which are close to the significance level $0.05$. In this case, we still reject the null hypothesis given the randomness of the p-value and conclude that both methods detect a statistically significant treatment effect in terms of the risk difference. 

\subsection{Model diagnosis \label{appen:real_diagnosis}}

In primary analysis and sensitivity analyses, we fit an MMRM from
a population-averaged perspective based on the observed data to get
a consistent estimator of the model parameter as $\hat{\theta}$.
The underlying assumptions for the MMRM are: (i) Normality: $Y_{i}:=(Y_{i1},\cdots,Y_{iT})^{\intercal}\mid(G_{i}=j,R_{iT}=1)\sim\mathcal{N}_{T}(\mu_{ij},\Sigma^{(j)})$;
(ii) The parameters controlling the missing mechanism and the model
parameters satisfy the separability condition, i.e., they are variation
independent. Note that the second assumption is not testable since
we cannot verify the missing mechanism only through observed components,
we focus on conducting a model diagnosis on the multivariate normality
of the observed components of the HAMD-17 trial data in each treatment
group.

Normal Q-Q plots for residuals from the marginal distribution of the
baseline response $Y_{1}$, and the full conditional distribution
of the following responses $Y_{2},\cdots,Y_{6}$ based on observed
data are given in Figure \ref{fig:diagnosis}. It supports the multivariate
normality assumption since the majority of residuals are within the
confidence region. We observe several outstanding residuals at the
last time point that mildly violate the normality assumption. However,
we still believe the normality holds in general.

\end{document}